\documentclass[english,twocolumn,prl,a4paper,nofootinbib,showkeys,showpacs]{revtex4}
\usepackage[T1]{fontenc}
\usepackage[latin9]{inputenc}
\usepackage{color}
\usepackage{babel}
\usepackage{prettyref}
\usepackage{mathrsfs}
\usepackage{amsmath}
\usepackage{amssymb}
\usepackage[unicode=true,
 bookmarks=true,bookmarksnumbered=false,bookmarksopen=false,
 breaklinks=true,pdfborder={0 0 1},backref=false,colorlinks=true]
 {hyperref}
\hypersetup{pdftitle={Transport equations for superconductors in the presence of spin interaction},
 pdfauthor={François Konschelle},
 pdfsubject={Abstract Quasi-classical theory of superconductivity provides a powerful and yet simple description of the superconductivity phenomenology. In particular, the Eilenberger and Usadel equations provide a neat simplification of the description of the superconducting state in the presence of disorder and electromagnetic interaction. However, the modern aspects of superconductivity require a correct description of the spin interaction as well. Here, we generalize the transport equations of superconductivity in order to take into account space-time dependent electromagnetic and spin interactions on equal footing. Using a gauge-covariant Wigner transformation for the Green-Gor'kov correlation functions, we establish the correspondence between the Dyson-Gor'kov equation and the quasi-classical transport equation in the time-dependent phase-space. We give the expressions for the gauge-covariant current and charge densities (quasi-particle, electric and spin) in the transport formulation. The generalized Eilenberger and Usadel limits of the transport equation are given, too. This study is devoted to the formal derivation of the equations of motion in the electromagnetic plus spin plus particle-hole space. The studies of some specific systems are postponed to future works. },
 pdfkeywords={Keywords: transport ; superconductivity ; spin-orbit ; spin-texture ; gauge ; quasi-classical ; theory ; Eilenberger ; Usadel},
 citecolor=blue, linkcolor=magenta, urlcolor=blue}
\usepackage{breakurl}

\makeatletter
\@ifundefined{textcolor}{}
{%
 \definecolor{BLACK}{gray}{0}
 \definecolor{WHITE}{gray}{1}
 \definecolor{RED}{rgb}{1,0,0}
 \definecolor{GREEN}{rgb}{0,1,0}
 \definecolor{BLUE}{rgb}{0,0,1}
 \definecolor{CYAN}{cmyk}{1,0,0,0}
 \definecolor{MAGENTA}{cmyk}{0,1,0,0}
 \definecolor{YELLOW}{cmyk}{0,0,1,0}
}

\makeatother

\begin{document}

\title{Transport equations for superconductors in the presence of spin interaction}

\author{Fran\c{c}ois Konschelle}

\affiliation{Institute for Quantum Information, RWTH Aachen University, 52056
Aachen, Germany}

\date{\today}


\begin{abstract}
Quasi-classical theory of superconductivity provides a powerful and
yet simple description of the superconductivity phenomenology. In
particular, the Eilenberger and Usadel equations provide a neat simplification
of the description of the superconducting state in the presence of
disorder and electromagnetic interaction. However, the modern aspects
of superconductivity require a correct description of the spin interaction
as well. Here, we generalize the transport equations of superconductivity
in order to take into account space-time dependent electromagnetic
and spin interactions on equal footing. Using a gauge-covariant Wigner
transformation for the Green-Gor'kov correlation functions, we establish
the correspondence between the Dyson-Gor'kov equation and the quasi-classical
transport equation in the time-dependent phase-space. We give the
expressions for the gauge-covariant current and charge densities (quasi-particle,
electric and spin) in the transport formulation. The generalized Eilenberger
and Usadel limits of the transport equation are given, too. This study
is devoted to the formal derivation of the equations of motion in
the electromagnetic plus spin plus particle-hole space. The studies
of some specific systems are postponed to future works. 
\end{abstract}
\maketitle
Without doubt, the theory of superconductivity, first established
by Bardeen, Cooper and Schrieffer \cite{bcs.1957_1,bcs.1957_2}, and
reformulated by Valatin \cite{valatin.1958}, Bogoliubov \cite{bogoliubov.1958},
Gor'kov \cite{gorkov.1958,b.abrikosov_gorkov} and Nambu \cite{nambu.1960,Nambu1960}
is a masterpiece of condensed matter in particular, and quantum field
theory in general. It consists in a few concepts -- a second-order
phase transition due to electron-phonon interaction, or a classical
gauge-symmetry breaking in high-energy language -- together with a
predictive power which provided breakthrough discoveries all along
the second half of the 20-th century. Among others, the BCS theory
and its close parent the Ginzburg-Landau model \cite{landau_ginzburg.1950,gorkov.1959}
allow the prediction of the vortex states \cite{Abrikosov1957a},
the Josephson effect \cite{josephson.1965}, the generation of massive
boson field at the phase transition \cite{Anderson1963,Englert1964,Higgs1964},
and the great family of the proximity effects \cite{b.de_gennes,b.tinkham},
... all experimentally well-established since.

The balance between a few concepts involved in a large number of novel
effects is certainly due to the robustness of the quasi-classical
description of superconductivity \cite{eilenberger.1968,larkin_ovchinnikov.1969,usadel.1970,Eliashberg1972}.
Indeed, most superconductors are characterized by a relevant energy
scale, namely the gap parameter energy, much smaller than the Fermi
energy. Then it becomes possible to adapt for superconductors the
quasi-classical theory developed for normal metals \cite{Kadanoff1962,Prange1964}.

Due to its success describing such vast problems as vortex in bulk,
Josephson and proximity effect in mesoscopic systems as well as the
competition between superconductivity and disorder, the quasi-classical
description of superconductivity was naturally extended to discuss
the competition between superconductor and magnetic orders. There,
the quasi-classical description opened a new era of discoveries, which
are too numerous to be listed here. We just mention that their possible
domain of applicability ranges from spintronic effects to some proposed
fundamental phases in neutron stars and in the early universe, passing
through original vortex states and new electronic devices based on
novel Josephson effects, see \textit{e.g.} \cite{golubov_kupriyanov.2004,casalbuoni.nardulli.2004,buzdin.2005_RMP,bergeret_volkov_efetov_R.2005}
and references therein. 

Whereas the first studies discussing the competing effect between
superconductivity and spin coupling focussed on constant ferromagnetic
field, there are emerging interests in the description of superconducting
systems having spin texture. The promises these systems carry on arose
several fields of research. On one side, there are fundamental questions
in bulk systems about the competitions between non-centrosymmetric
magnetic order and the superconducting phase, leading to interesting
magneto-electric effects, original vortex lattices, helical superconducting
phase, ... \cite{Bauer2012a}. On the other side, the presence of
spin-orbit interaction in superconducting wires has been predicted
to generate topological states of matter, possibly helpful for quantum
computation \cite{Alicea2012,Beenakker2011}. In these wires hosting
Majorana modes, there are still vivid discussions about the role of
impurities, the nature of the competition between the proximity effect
and the spin-texture, ... see \textit{e.g.} \cite{Franz2013}. Moreover,
having ferromagnetic plus spin-orbit coupling in a superconducting
wire does not seem to be rich enough to provide universal quantum
computation, and people are recently discussing spin-texture in quantum-Hall
plus superconducting heterostructures in order to generate possible
parafermions \cite{Refael2012,Clarke2013a,Clarke2013}.

A quasi-classical description of superconductivity able to take into
account spin-texture and impurities is thus highly desirable. Of course,
it exists several ways to perturbatively simplify this complicated
problem, as discussing diffusive systems, or perturbatively weak disorder,
and/or small spin-orbit effect for Majorana wires for instance \cite{Brouwer2011,Pientka2012,Neven2013}.
Note also the literature associated to the inclusion of spin-orbit
effect in bulk superconductor or superfluid without exchange field
\cite{Schopohl1980,Hayashi2006,Agterberg2006,Sauls2009}, or the alternative
possibility to use topological superconductivity ($p$-wave) as an
effective model for spin-textured superconductor \cite{Stanev2014}%
\footnote{Note also that, after completion of this work, I became aware of a
similar study by Bergeret and Tokatly, which use the same method as
the one in this paper to obtain similar equations \cite{Bergeret2014}.
A discussion of the main differences between their paper and mine
can be find in \prettyref{sec:Potential-or-gauge-potential}.%
}. Nevertheless, a reliable construction of a general quasi-classical
theory should be of interest in several active research fields. This
is also the case in normal and semiconducting diffusive systems including
spin-texture. There, it has been shown recently that a gauge-theory
construction provides a transparent procedure for the derivation of
transport equations \cite{Gorini2010} including spin effects.

This paper is devoted to the question of the inclusion of the spin
texture (Zeeman plus spin-orbit interaction say) in the superconductivity
phenomenology. Here, we recognize the venerable principle of gauge
redundancy (see \textit{e.g.} \cite{Frohlich1993}) as a fruitful
tool for the construction of a transport theory of superconductivity,
including space-time dependent spin and charge fields. In particular,
we generalize the results from \cite{Gorini2010} in order to include
the superconducting correlation functions. We adapt the description
of the quark-gluon plasma \cite{Elze1986a} to the non-relativistic
situation of a superconductor in the presence of some generic Abelian
(electromagnetic for instance) and non-Abelian (spin and particle-hole)
gauge-fields. 

I aim this paper to be as pedagogical as possible, especially in the
sometimes confusing adoption of the mixed-Fourier transformation \cite{Eckern1981,serene_rainer.1983,Rammer1986,belzig_wilhem.1999,b.kopnin},
which is nothing more than a Wigner transformation \cite{Hillery1984,Polkovnikov2009},
here fruitfully made gauge-covariant, see \prettyref{sec:Transport-equations}.
To that purpose, I sum-up the conventions I follow in \prettyref{sec:Green-Gorkov-equation}
and \prettyref{sec:Definition-and-Convention}, and I provide explicit
-- though lengthy -- calculations in an appendix. Even though the
calculation of the appendix can be generalized straightforwardly to
higher orders, I discuss in the main text the explicit model of non-relativistic
free particles in the quasi-classical limit. This limits the present
study to the Rashba-like spin-orbit effect when the spin interaction
is linear in momentum. 

I generalize the BCS treatment given by Gor'kov \cite{gorkov.1958}
to the non-Abelian gauge theory in \prettyref{sec:Green-Gorkov-equation}.
I then discuss the equations of motion for the gauge fields in \prettyref{sec:Definition-and-Convention},
following the standard treatment \cite{Itzykson2006}. The transport
equations at the quasi-classical level are given in \prettyref{sec:Transport-equations}.
There I establish the main results of this paper, namely expressions
\prettyref{eq:transport-diff} and \prettyref{eq:transport-sum}.
Then I turn to the Eilenberger (\prettyref{sec:Eilenberger-equation},
eq.\prettyref{eq:Eilenberger}) and Usadel (\prettyref{sec:Usadel-equation},
eq.\prettyref{eq:Usadel}) limits of these equations, when the relevant
energies are constrained to the proximity of the Fermi energy in the
general and diffusive limit, respectively. Especially, the derivation
from the transport equation to the Eilenberger one is treated in full
details as well as the so-called normalization condition (\prettyref{sec:Eilenberger-equation}).
The two last sections sum up an alternative derivation of the gauge-covariant
Eilenberger equation (\prettyref{sec:Poor-man-Eilenberger}), perhaps
more comprehensible than the lengthy calculation of \prettyref{sec:Eilenberger-equation},
and a discussion of the usual treatment of a constant exchange field
in the quasi-classical limit (\prettyref{sec:Potential-or-gauge-potential}).
Some perspectives of the present work are given alongside the conclusion
in \prettyref{sec:Conclusion-and-Perspectives}.

\section{Matter field: Dyson-Gor'kov's equations\label{sec:Green-Gorkov-equation}}

We start our discussion with a brief summary of some known results
in the theory of superconductivity. In fact, this work starts from
the Dyson-Gor'kov equations at zero temperature \cite{gorkov.1958,b.abrikosov_gorkov}.
These equations represent the evolution of the quantum-field correlation
functions in space-time. According to the Gor'kov theory, the superconducting
systems are described in the so-called Nambu space, or particle-hole
space. Here, I generalize the Gor'kov theory toward a non-Abelian
theory including both the Nambu and the spin space, in addition to
the usual Abelian electromagnetic space. Reader familiar with the
quantum field theory of superconductivity can skip this section up
to the equation \prettyref{eq:Delta}, reader also familiar with the
concept of gauge theory can skip this section entirely, as well as
\prettyref{sec:Definition-and-Convention}. 

Since we will discuss gauge properties in space-time, it is convenient
to use the relativistic quadri-vectors notations. They are defined
as $x^{\mu}=\left(ct,\boldsymbol{x}\right)$ and $\partial_{\mu}\equiv\left(\partial_{ct},\boldsymbol{\partial_{x}}\right)$
with the metric tensor $g_{\mu\nu}=\left(1,-1,-1,-1\right)$. Later
on we will define an energy-momentum 4-vector $p^{\mu}=\left(E/c,\boldsymbol{p}\right)$
with $E=\hslash\omega$ which defines the angular frequency $\omega$,
and $\boldsymbol{p}=\hbar\boldsymbol{k}$. Contracted indices are
implicitly summed, the greek ones being over the full space-time $\mu,\nu\equiv\left(0,1,2,3\right)$,
whereas the latin ones are only over the space variables $i,j,k=\left(1,2,3\right)$.
To not confound the indices with the imaginary unit vector, the latter
is noted in bold $\mathbf{i}^{2}=-1$. As much as possible, I try
to avoid using bold symbols for the collection of the components of
the vectors, which are preferably written in terms of their components.
The bold letters are kept for the symbolic notation of the elements
in the Nambu space. When not possible otherwise, I use italic bold
letters to describe vectors in space-time. I use also the notation
for the central dot to represent the scalar product, either in space-time
or in space: for instance $p\cdot x/\hbar=p_{\mu}x^{\mu}/\hbar=\omega t-\boldsymbol{k\cdot x}=\omega t+k^{i}x_{i}$.

In order to discuss the quantum field theory of superconductivity,
I adopt the spinor notation in the Nambu space
\begin{equation}
\Psi\left(x\right)=\left(\begin{array}{c}
\Psi_{\uparrow}\left(x\right)\\
\Psi_{\downarrow}\left(x\right)
\end{array}\right)\;\text{and}\;\tilde{\Psi}\left(x\right)=\left(\begin{array}{cc}
\Psi_{\uparrow}\left(x\right) & \Psi_{\downarrow}\left(x\right)\end{array}\right)\label{eq:Psi-spin}
\end{equation}
with the convention $\Psi^{\dagger}=\tilde{\Psi}^{\ast}$, where $\Psi\left(x\right)$
annihilates a fermion at space-time position $x$, whereas $\Psi^{\dagger}\left(x\right)$
creates a fermion at that position. The Green-Gor'kov correlation
functions in space-time are defined via a matrix in the Nambu space
\begin{align}
\mathbf{G} & =\dfrac{\mathbf{i}}{\hbar}\left\langle \hat{T}\left[\left(\begin{array}{c}
-\Psi\left(x_{1}\right)\\
\tilde{\Psi}^{\dagger}\left(x_{1}\right)
\end{array}\right)\otimes\left(\begin{array}{cc}
\Psi^{\dagger}\left(x_{2}\right) & \tilde{\Psi}\left(x_{2}\right)\end{array}\right)\right]\right\rangle \nonumber \\
 & =\left(\begin{array}{cc}
G\left(x_{1},x_{2}\right) & -F\left(x_{1},x_{2}\right)\\
F^{\dagger}\left(x_{1},x_{2}\right) & G^{\dagger}\left(x_{1},x_{2}\right)
\end{array}\right)\label{eq:G-matrix-def}
\end{align}
where $x_{1,2}\equiv x_{1,2}^{\mu}$, the $\hat{T}$ operator is the
time-ordering operator, and the average $\left\langle \cdots\right\rangle $
is a quantum average \cite{b.abrikosov_gorkov}. The definition \prettyref{eq:G-matrix-def}
in terms of a tensor product will simplify the gauge transformation
treatment below. Note that we do not describe further the sub-space
for the functions $G$, $F$, ... Additionally, the spin and charge
structure will be entirely defined through the gauge-potentials defined
later. In contrary, the Pauli matrices notation will be of importance.
I use the $\tau_{i}$ matrices to represent the Nambu algebra, and
the $\sigma_{i}$ matrices to represent the spin algebra.

We start from the simplest model of a free electron gas interacting
through the usual BCS interaction and described by the Hamiltonian
$H=H_{0}+H_{\text{int}}$ with
\begin{equation}
H_{0}=\int dx\left[\Psi^{\dagger}\left(x\right)\left(-\dfrac{\hslash^{2}}{2m}\boldsymbol{\partial_{x}\cdot\partial_{x}}-\mu\right)\Psi\left(x\right)\right]\label{eq:H-0}
\end{equation}
where $\mu$ is the chemical potential, and 
\begin{multline}
H_{\text{int}}=\int dx\dfrac{V_{0}\left(x\right)}{2}\times\\
\left[\tilde{\Psi}\left(x\right)\mathbf{i}\sigma_{2}\Psi\left(x\right)\right]^{\dagger}\left[\tilde{\Psi}\left(x\right)\mathbf{i}\sigma_{2}\Psi\left(x\right)\right]\label{eq:H-int}
\end{multline}
is a two-body interaction, with the spin-independent $V_{0}$ being
an attractive interaction strength in the superconducting regions
of space, and otherwise vanishing. The Heisenberg equation of motion\textbf{
$\mathbf{i}\hbar\partial_{t}\Psi=\left[\Psi,H\right]$} leads to 
\begin{equation}
\int dy\left[\mathbf{G}^{-1}\left(x_{1},y\right)\mathbf{G}\left(y,x_{2}\right)\right]=\delta\left(x_{1}-x_{2}\right)\label{eq:Dyson-1}
\end{equation}
with $\mathbf{G}^{-1}\left(x_{1},y\right)=\mathbf{G}_{0}^{-1}\left(x_{1}\right)\delta\left(x_{1}-y\right)$
and 
\begin{equation}
\mathbf{G}_{0}^{-1}\left(x\right)=\dfrac{\hslash^{2}}{2m}\boldsymbol{\partial_{x}\cdot\partial_{x}}+\mu+\left(\begin{array}{cc}
\mathbf{i}\hbar\partial_{t} & \Delta\left(x\right)\\
-\Delta^{\dagger}\left(x\right) & -\mathbf{i}\hbar\partial_{t}
\end{array}\right)\label{eq:G0}
\end{equation}
is the so-called propagator. The gap parameter 
\begin{equation}
\Delta\left(x\right)=V_{0}\left(x\right)\left\langle \hat{T}\left[\tilde{\Psi}\left(x\right)\left(\mathbf{i}\sigma_{2}\right)\Psi\left(x\right)\right]\right\rangle \left(\mathbf{i}\sigma_{2}\right)^{\dagger}\label{eq:Delta}
\end{equation}
is defined self-consistently as 
\begin{equation}
\Delta_{0}\left(x_{2}\right)=-\mathbf{i}\hbar\lim_{x_{1}\rightarrow x_{2}}V_{0}\left(x_{1}\right)\text{Tr}\left\{ \mathbf{i}\sigma_{2}F\left(x_{1},x_{2}\right)\right\} \label{eq:self-consistent}
\end{equation}
 with $\Delta\left(x\right)=\Delta_{0}\left(x\right)\left(\mathbf{i}\sigma_{2}\right)^{\dagger}$
and the trace is taken over the sub-space(s) of the $F\left(x_{1},x_{2}\right)$
matrix. The gap parameter appears in \prettyref{eq:Dyson-1} thanks
to a mean-field decoupling in the Cooper pairing channel, see \cite{b.abrikosov_gorkov}
for more details.

It is noteworthy to realize that the interaction Hamiltonian $H_{\text{int}}$
is both $U\left(1\right)$ and $SU\left(2\right)$ gauge invariant,
\textit{i.e.} it is invariant under the transformation%
\footnote{In this paper, I propose to use the symbol $\rightsquigarrow$ to
denote the gauge transformation mapping. No confusion with its definition
in symbolic calculation is possible.%
}
\begin{align}
\Psi\left(x\right) & \rightsquigarrow R\left(x\right)\Psi\left(x\right)\nonumber \\
\Psi^{\dagger}\left(x\right) & \rightsquigarrow\Psi^{\dagger}\left(x\right)R^{\dagger}\left(x\right)\label{eq:gauge-transform-Psi}
\end{align}
with $R\left(x\right)\in U\left(1\right)\otimes SU\left(2\right)$.
Since it describes singlet spin coupling, any spin rotation will let
it unaffected hence it is $SU\left(2\right)$ gauge invariant -- mathematically
speaking this corresponds to the remark that $e^{\mathbf{i}n^{i}\sigma_{i}}\sigma_{2}e^{\left(\mathbf{i}n^{i}\sigma_{i}\right)^{\ast}}=\sigma_{2}$
for any unit vector components $n^{i}$. Since it has the same number
of $\Psi\left(x\right)$ as $\Psi^{\dagger}\left(x\right)$, it is
$U\left(1\right)$ gauge invariant. We also realize that $H_{\text{int}}$
is nothing but the usual $s$-wave interaction Hamiltonian \cite{b.abrikosov_gorkov,b.de_gennes,b.tinkham}.
Then we could promote the equation of motion \prettyref{eq:Dyson-1}
to be $U\left(1\right)\otimes SU\left(2\right)$ gauge covariant in
principle. 

From its definition \prettyref{eq:G-matrix-def}, a gauge transformation
\prettyref{eq:gauge-transform-Psi} of the Green-Gor'kov matrix $\mathbf{G}$
reads 
\begin{equation}
\mathbf{G}\left(x_{1},x_{2}\right)\rightsquigarrow\mathbf{R}\left(x_{1}\right)\mathbf{G}\left(x_{1},x_{2}\right)\mathbf{R}^{-1}\left(x_{2}\right)\label{eq:gauge-transform-Gxx}
\end{equation}
with $\mathbf{R}$ given by 
\begin{equation}
\mathbf{R}\left(x\right)=\left(\begin{array}{cc}
R\left(x\right) & 0\\
0 & R^{\ast}\left(x\right)
\end{array}\right)\label{eq:U}
\end{equation}
in the particle-hole space. Since we consider some unitary matrices
$R^{\dagger}R=1$, one has $\mathbf{R}^{-1}=\mathbf{R}^{\dagger}$.
Note nevertheless that the left and right $\mathbf{R}$ transformation
matrices are not evaluated at the same point: the Green-Gor'kov functions
are two-points correlation functions in space-time. Then the general
covariance of the Green-Gor'kov equations is constructed under the
demand that the transformation
\begin{multline}
\mathbf{G}_{0}^{-1}\left(x_{1}\right)\mathbf{G}\left(x_{1},x_{2}\right)\rightsquigarrow\\
\mathbf{R}\left(x_{1}\right)\mathbf{G}_{0}^{-1}\left(x_{1}\right)\mathbf{G}\left(x_{1},x_{2}\right)\mathbf{R}^{\dagger}\left(x_{2}\right)
\end{multline}
works after a proper substitution of the derivatives with some covariant
derivatives, the usual minimal or Weyl's substitution \cite{Landau1974}.
One verifies easily that the correct minimal substitution reads
\begin{equation}
\mathbf{G}_{0}^{-1}\left(x\right)=\mathbf{i}\hbar c\tau_{3}\mathbf{D}_{0}\left(x\right)-\dfrac{\hbar^{2}}{2m}\mathbf{D}^{i}\mathbf{D}_{i}\left(x\right)+\boldsymbol{\Delta}\left(x\right)\label{eq:G0-covariant}
\end{equation}
with the covariant derivative 
\begin{align}
\mathbf{D}_{\mu}\left(x\right) & =\dfrac{\partial}{\partial x^{\mu}}+\mathbf{i}\tau_{3}\left(\begin{array}{cc}
A_{\mu} & 0\\
0 & A_{\mu}^{\ast}
\end{array}\right)=\partial_{\mu}+\mathbf{i}\mathbf{A}_{\mu}\left(x\right)\label{eq:cov-deriv-super}
\end{align}
defining the gauge potential $\mathbf{A}_{\mu}$. It transforms according
to
\begin{equation}
\mathbf{A}_{\mu}\left(x\right)\rightsquigarrow\mathbf{R}\left(x\right)\mathbf{A}_{\mu}\left(x\right)\mathbf{R}^{\dagger}\left(x\right)-\mathbf{i}\mathbf{R}\left(x\right)\partial_{\mu}\mathbf{R}^{\dagger}\left(x\right)\label{eq:gauge-transform-A}
\end{equation}
when the Green-Gor'kov matrix transforms as \prettyref{eq:gauge-transform-Gxx}.
One associates the gauge field
\begin{equation}
\mathbf{F}_{\mu\nu}\left(x\right)=\partial_{\mu}\mathbf{A}_{\nu}-\partial_{\nu}\mathbf{A}_{\mu}+\mathbf{i}\left[\mathbf{A}_{\mu}\left(x\right),\mathbf{A}_{\nu}\left(x\right)\right]\label{eq:F-bold}
\end{equation}
with the gauge potential. It transforms covariantly as well 
\begin{equation}
\mathbf{F}_{\mu\nu}\left(x\right)\rightsquigarrow\mathbf{R}\left(x\right)\mathbf{F}_{\mu\nu}\left(x\right)\mathbf{R}^{\dagger}\left(x\right)\label{eq:gauge-transform-F}
\end{equation}
 under the gauge transformation \prettyref{eq:gauge-transform-A}.
We also define
\begin{equation}
\boldsymbol{\Delta}\left(x\right)=\tau_{+}\Delta\left(x\right)-\tau_{-}\Delta^{\dagger}\left(x\right)\label{eq:Delta-bold}
\end{equation}
for the gap-parameter matrix in the Nambu space, with $\tau_{\pm}=\left(\tau_{1}\pm\mathbf{i}\tau_{2}\right)/2$.
We remark that we could have included the gap parameter in the $\mathbf{D}_{0}$,
but there are difficulties with dealing with non-diagonal covariant
derivatives in the particle-hole space, see the end of \prettyref{sec:Transport-equations}
for more details. Note also that the gap parameter is affected by
the gauge transformation as 
\begin{equation}
\boldsymbol{\Delta}\left(x\right)\rightsquigarrow\mathbf{R}\left(x\right)\boldsymbol{\Delta}\left(x\right)\mathbf{R}^{\dagger}\left(x\right)\label{eq:gauge-transform-gap}
\end{equation}
though there is no signature of this in the propagator, since its
gauge transformation is absorbed by the correct Green-Gor'kov correlation
function when writing the $\mathbf{G}_{0}^{-1}\left(x_{1}\right)\mathbf{G}\left(x_{1},x_{2}\right)$
product explicitly. According to our general prescription, we do not
write explicit expressions for the gauge fields $A_{\mu}$ for the
moment. The impatient reader who wants to know why the chemical potential
disappeared in \prettyref{eq:G0-covariant} can check \prettyref{eq:A-def}.

A similar calculation for the equation of motion 
\begin{equation}
\int dy\left[\mathbf{G}\left(x_{1},y\right)\left[\mathbf{G}^{-1}\left(y,x_{2}\right)\right]^{\dagger}\right]=\delta\left(x_{1}-x_{2}\right)\label{eq:Dyson-2}
\end{equation}
gives the same propagator \prettyref{eq:G0-covariant} now in its
adjoint form $\left[\mathbf{G}^{-1}\left(y,x_{2}\right)\right]^{\dagger}=\left[\mathbf{G}_{0}^{-1}\left(x_{2}\right)\right]^{\dagger}\delta\left(y-x_{2}\right)$.
One verifies that
\begin{multline}
\mathbf{G}\left(x_{1},x_{2}\right)\mathbf{G}_{0}^{-1}\left(x_{2}\right)\rightsquigarrow\\
\mathbf{R}\left(x_{1}\right)\mathbf{G}\left(x_{1},x_{2}\right)\left[\mathbf{G}_{0}^{-1}\left(x_{2}\right)\right]^{\dagger}\mathbf{R}^{\dagger}\left(x_{2}\right)
\end{multline}
under a gauge transformation. Note that $\mathbf{D}_{\mu}^{\dagger}=\partial_{\mu}-\mathbf{i}\mathbf{A}_{\mu}$
where the derivative applies to the left and the gauge potential is
supposed Hermitian. When discussing non-Abelian gauge, the gauge-potential
$A_{\mu}$ in \prettyref{eq:cov-deriv-super} does not commute with
the Green-Gor'kov functions. 

We conclude this section with a few words about impurities. They are
usually accounted for in a self-energy term, which corresponds to
defining \cite{b.abrikosov_gorkov,b.kopnin,Gorkov2008}
\begin{align}
\mathbf{G}^{-1}\left(x_{1},x_{2}\right) & =\mathbf{G}_{0}^{-1}\left(x_{1}\right)\delta\left(x_{1}-x_{2}\right)-\boldsymbol{\Sigma}\left(x_{1},x_{2}\right)\label{eq:G-1-self-energy}
\end{align}
in the equations \prettyref{eq:Dyson-1} and \prettyref{eq:Dyson-2},
respectively. 

Equations \prettyref{eq:Dyson-1} and \prettyref{eq:Dyson-2} are
the equations of motion for the Green-Gor'kov correlators in the presence
of space-time dependent non-Abelian gauge-field and impurities realization.
For the moment we did not precise the gauge structure explicitly.
This allow the equations \prettyref{eq:Dyson-1} and \prettyref{eq:Dyson-2}
to be of full generality. Note that the corresponding Bogoliubov-de
Gennes \cite{b.de_gennes} and Landau-Ginzburg \cite{landau_ginzburg.1950}
formalisms can be adapted as well to the non-Abelian gauge interaction:
it is sufficient to replace the Abelian covariant derivative with
the non-Abelian one in \prettyref{eq:cov-deriv-super}. We stress
one more time that the generalization toward a non-Abelian formalism
was here possible thanks to the invariance of the interaction Hamiltonian
\prettyref{eq:H-int} under a gauge transformation. In the next section
we establish the equations of motion for the gauge fields, and connect
them to the Green-Gor'kov functions.

\section{Gauge field: Maxwell's and Yang-Mills's equations\label{sec:Definition-and-Convention}}

Since the concept of the non-Abelian gauge theory could be new for
a few condensed matter physicists, I discuss it in this section. The
general form of the gauge potential and the gauge field for the $U\left(1\right)\otimes SU\left(2\right)\otimes SU\left(2\right)$
gauge redundancy in the Nambu $\otimes$ spin $\otimes$ electromagnetic
space is discussed, as well as the associated equations of motion.
In particular, the gauge formalism allows to define the charge and
current densities associated with the different gauge fields. This
section ends up with a discussion of the explicit form of the gauge
potential in non-trivial situations, namely when Zeeman and/or spin-orbit
interactions are participating to the electron dynamics. Reader familiar
with the principles of gauge theory can skip this section.

From the discussion of the above section, one can establish an effective
classical Lagrangian density 
\begin{align}
L_{\psi} & =\boldsymbol{\psi}^{\dagger}\left[\mathbf{i}\hbar c\mathbf{D}_{0}+\dfrac{\hbar^{2}}{2m}\mathbf{D}_{j}\mathbf{D}_{j}+\boldsymbol{\Delta}\right]\boldsymbol{\psi}
\end{align}
where $\boldsymbol{\psi}^{\dagger}=\left(\begin{array}{cc}
\psi^{\dagger} & \tilde{\psi}\end{array}\right)$ is a classical spinor in the Nambu space, and eventually $\psi$,
its transpose $\tilde{\psi}$ and its adjoint $\psi^{\dagger}$ are
spinors themselves. In the following they will indeed be spinors in
the spin-space. Whether the $\boldsymbol{\psi}$ classical spinor
field makes sense in the Nambu space or not is of no concern to us:
our primary interest is in the construction of the currents in this
section. By construction, $L_{\psi}$ is invariant under the gauge
transform
\begin{equation}
\boldsymbol{\psi}\left(x\right)\rightsquigarrow\mathbf{R}\left(x\right)\boldsymbol{\psi}\left(x\right)\;\text{and}\;\boldsymbol{\psi}^{\dagger}\left(x\right)\rightsquigarrow\boldsymbol{\psi}^{\dagger}\left(x\right)\mathbf{R}^{\dagger}\left(x\right)
\end{equation}
for the classical spinor, and \prettyref{eq:gauge-transform-A}, \prettyref{eq:gauge-transform-F}
and \prettyref{eq:gauge-transform-gap} for the gauge potential and
field, and the gap matrix $\boldsymbol{\Delta}$ transformations.

The variation of the action $S_{\psi}=\int dx\left[L_{\psi}\right]$
with respect to the matter-field $\boldsymbol{\psi}$ gives the classical
equations of motion $\mathbf{G}_{0}^{-1}\left(x\right)\boldsymbol{\psi}\left(x\right)=0$
and its adjoint, with $\mathbf{G}_{0}^{-1}$ in \prettyref{eq:G0}.
Next, the variation of the matter-field action $S_{\psi}$ with respect
to the gauge potential gives some current and charge densities. Since
the gauge potential $\mathbf{A}_{\mu}$ is a scalar in the particle-hole
space, one has 
\begin{equation}
\delta S_{\psi}=\int dx\left[-\dfrac{c}{\hbar}\rho_{n}\delta\mathbf{A}_{0}-\dfrac{j_{n}^{i}}{\hbar}\delta\mathbf{A}_{i}\right]
\end{equation}
(plus the variations with respect to $\boldsymbol{\psi}$ and $\boldsymbol{\psi}^{\dagger}$
which are discarded here for commodity) with 
\begin{equation}
\rho_{n}\left(x\right)=\boldsymbol{\psi}^{\dagger}\left(x\right)\boldsymbol{\psi}\left(x\right)
\end{equation}
for the particle density and 
\begin{equation}
j_{n}^{i}\left(x\right)=\dfrac{\mathbf{i}\hbar}{2m}\boldsymbol{\psi}^{\dagger}\left(x\right)\left[\mathbf{D}_{j}^{\dagger}\left(x\right)-\mathbf{D}_{j}\left(x\right)\right]\boldsymbol{\psi}\left(x\right)
\end{equation}
for the particle current density. This current is neutral and conserved:
$\partial_{t}\rho_{n}+\boldsymbol{\partial_{x}\cdot j}_{n}=0$.

To find the microscopic spin and electric currents, we expand the
gauge potential thanks to the representation
\begin{equation}
\mathbf{A}_{\mu}\equiv\left(\dfrac{e}{\hbar c}\varphi\tau_{3}-\dfrac{\mu}{\hbar c}+\dfrac{g}{\hbar c}a_{0}^{i}\dfrac{\mathbf{s}_{i}}{2},-\dfrac{e}{\hbar}A_{j}\tau_{3}-\dfrac{g}{\hbar}a_{j}^{i}\dfrac{\mathbf{s}_{i}}{2}\right)\label{eq:A-def}
\end{equation}
where we define 
\begin{equation}
\mathbf{s}_{i}=\left(\begin{array}{cc}
\sigma_{i} & 0\\
0 & -\sigma_{i}^{\ast}
\end{array}\right)
\end{equation}
the spin matrix in the Nambu space. All the components $\varphi$,
$A_{j}$ and $a_{\mu}^{i}$ are real. We associate the charges $e$
and $g$ to the electric $U\left(1\right)$ and the spin $SU\left(2\right)$
gauge fields, respectively, and $\mu$ is the chemical potential.
One then has
\begin{equation}
\delta S_{\psi}=\int dx\left[-j_{e}^{i}\delta A_{i}-j_{s}^{ik}\delta a_{i}^{k}-\rho_{e}\delta\varphi-\rho_{s}^{k}\delta a_{0}^{k}\right]
\end{equation}
with the electric charge and current densities
\begin{equation}
\rho_{e}=e\boldsymbol{\psi}^{\dagger}\tau_{3}\boldsymbol{\psi}\;\text{and}\; j_{e}^{i}=\dfrac{\mathbf{i}\hbar e}{2m}\boldsymbol{\psi}^{\dagger}\left[\mathbf{D}_{i}\tau_{3}-\tau_{3}\mathbf{D}_{i}^{\dagger}\right]\boldsymbol{\psi}
\end{equation}
and the spin charge and current densities
\begin{equation}
\rho_{s}^{k}=\dfrac{g}{2}\boldsymbol{\psi}^{\dagger}\mathbf{s}_{k}\boldsymbol{\psi}\;\text{and}\; j_{s}^{ik}=\dfrac{\mathbf{i}\hbar g}{4m}\boldsymbol{\psi}^{\dagger}\left[\mathbf{D}_{i}\mathbf{s}^{k}-\mathbf{s}^{k}\mathbf{D}_{i}^{\dagger}\right]\boldsymbol{\psi}
\end{equation}
in term of the classical spinor $\boldsymbol{\psi}$. One can nevertheless
promote the above expressions to operators formula. It consists in
promoting the spinor $\boldsymbol{\psi}$ to the second quantized
fields $\boldsymbol{\Psi}$ and to average the above expressions for
the densities to obtain
\[
\rho_{n}=-\mathbf{i}\hbar\lim_{x_{1}\rightarrow x_{2}}\text{Tr}\left\{ \tau_{3}\mathbf{G}\left(x_{1},x_{2}+0\right)\right\} 
\]
\begin{multline}
j_{n}^{i}=\dfrac{\hbar^{2}}{2m}\lim_{x_{1}\rightarrow x_{2}}\text{Tr}\left\{ \mathbf{D}_{i}\left(x_{1}\right)\tau_{3}\mathbf{G}\left(x_{1},x_{2}+0\right)-\right.\\
\left.\tau_{3}\mathbf{G}\left(x_{1},x_{2}+0\right)\mathbf{D}_{i}^{\dagger}\left(x_{2}\right)\right\} \label{eq:neutral}
\end{multline}
\[
\rho_{e}=-\mathbf{i}e\hbar\lim_{x_{1}\rightarrow x_{2}}\text{Tr}\left\{ \mathbf{G}\left(x_{1},x_{2}+0\right)\right\} 
\]
\begin{multline}
j_{e}^{i}=\dfrac{e\hbar^{2}}{2m}\lim_{x_{1}\rightarrow x_{2}}\text{Tr}\left\{ \mathbf{D}_{j}\left(x_{1}\right)\mathbf{G}\left(x_{1},x_{2}+0\right)-\right.\\
\left.\mathbf{G}\left(x_{1},x_{2}+0\right)\mathbf{D}_{j}^{\dagger}\left(x_{2}\right)\right\} \label{eq:electric}
\end{multline}
\[
\rho_{s}^{k}=-\mathbf{i}\dfrac{g}{2}\hbar\lim_{x_{1}\rightarrow x_{2}}\left[\text{Tr}\left\{ \tau_{3}\mathbf{s}_{k}\mathbf{G}\left(x_{1},x_{2}+0\right)\right\} \right]
\]
\begin{multline}
j_{s}^{ik}=\dfrac{g\hbar^{2}}{4m}\lim_{x_{1}\rightarrow x_{2}}\text{Tr}\left\{ \mathbf{D}_{i}\left(x_{1}\right)\tau_{3}\mathbf{s}_{k}\mathbf{G}\left(x_{1},x_{2}+0\right)\right.-\\
\left.\tau_{3}\mathbf{s}_{k}\mathbf{G}\left(x_{1},x_{2}+0\right)\mathbf{D}_{i}^{\dagger}\left(x_{2}\right)\right\} \label{eq:spin}
\end{multline}
in term of the Green functions \prettyref{eq:G-matrix-def}. We did
not introduce a new notation for the averaged densities, since we
will only use the expressions \prettyref{eq:neutral}, \prettyref{eq:electric}
and \prettyref{eq:spin} for the neutral particle and current densities,
and the electric and spin charge and current densities, respectively.

Note that both the electric \prettyref{eq:electric} and spin \prettyref{eq:spin}
currents contain some magneto-electric contributions proportional
to $ge$. This feature is a hallmark of the non-relativistic gauge
theory we discuss in this section%
\footnote{More precisely this is the result of the second order space (covariant)
derivative in the non-relativistic model of \prettyref{sec:Green-Gorkov-equation}.%
}, and opens some interesting perspectives in the manipulation of the
quantum state via coherent circuits, as well as in the electromagnetic
response in spin textured superconductors \cite{Bauer2012a}.

An other important property of the gauge theory is its ability to
provide the equations of motion for the gauge field itself. To establish
them, we have to complement $L_{\psi}$ with a Lagrangian density
for the gauge field. Thus we represent the gauge fields as 
\begin{equation}
\mathbf{F}_{\mu\nu}=\dfrac{e}{\hbar}F_{\mu\nu}+\dfrac{g}{\hbar}F_{\mu\nu}^{k}\dfrac{\sigma_{k}}{2}\label{eq:F-complete}
\end{equation}
which we inject in the definition \prettyref{eq:F-bold} to get 
\begin{equation}
F_{\mu\nu}=\partial_{\mu}A_{\nu}-\partial_{\nu}A_{\mu}\label{eq:F-charge-def}
\end{equation}
for the gauge field in the charge sector, when $A_{\mu}\equiv\left(\varphi/c,-\boldsymbol{A}\right)$
and 
\begin{equation}
F_{\mu\nu}^{k}=\partial_{\mu}a_{\nu}^{k}-\partial_{\nu}a_{\mu}^{k}-\dfrac{g}{\hbar}\varepsilon_{ijk}a_{\mu}^{i}a_{\nu}^{j}\label{eq:F-spin-def}
\end{equation}
in the spin sector, with the gauge potential components as in \prettyref{eq:A-def},
and $\varepsilon_{ijk}$ the complete antisymmetric symbol. Then the
Lagrangian density 
\begin{equation}
L_{F}=-\dfrac{1}{4\mu_{0}}F_{\mu\nu}F_{\mu\nu}-\dfrac{1}{4}F_{\mu\nu}^{k}F_{\mu\nu}^{k}\label{eq:F-def}
\end{equation}
is gauge invariant under the transformation \prettyref{eq:gauge-transform-A}.
This can be easily verified by noting that \prettyref{eq:F-def} can
be written as some traces of \prettyref{eq:F-complete} and using
the transformation law \prettyref{eq:gauge-transform-F}, for more
details see \cite{Itzykson2006}. Next the variation of the total
action $S=\int dx\left[L_{\psi}+L_{F}\right]$ with respect to the
gauge-potential gives the usual Maxwell's equations \cite{Itzykson2006}
\begin{equation}
\partial_{\mu}F_{\mu\nu}=\mu_{0}J_{\nu}\label{eq:motion-F-Ab}
\end{equation}
and the so-called Yang-Mills's equations \cite{Yang1954} 
\begin{equation}
\partial_{\mu}F_{\mu\nu}^{k}-\dfrac{g}{\hbar}\varepsilon_{ijk}a_{\mu}^{i}F_{\mu\nu}^{j}=J_{\nu}^{k}\label{eq:motion-F-NAb}
\end{equation}
for the equation of motion of the gauge-fields, with $J^{\mu}\equiv\left(c\rho_{e},\boldsymbol{j}_{e}\right)$
and $J_{k}^{\mu}\equiv\left(c\rho_{s}^{k},\boldsymbol{j}_{s}^{k}\right)$
the quadri-currents for charge and spin, respectively. Note that the
equations of motion \prettyref{eq:motion-F-NAb} are non-linear in
term of the gauge potential, due to \prettyref{eq:F-spin-def}.

To illustrate the gauge formalism, let me give some examples of the
gauge-potentials in simple systems. Obviously when there is no spin
interaction, there is no need for a non-Abelian gauge potential. Yet
injecting the Abelian potential as above is the usual way to generate
the interaction between electric charge and fields, at both the classical
and quantum level \cite{Itzykson2006}. Next, a Zeeman effect usually
appears as the additional $L_{\text{Z}}=-h^{i}\left(x\right)\sigma_{i}$
term in the Lagrangian density for the otherwise free particles. Then
it can be absorbed as a gauge-potential $a_{0}^{i}=h^{i}\left(x\right)$
with $g=2/\hbar c$ and all the other gauge-potentials are zero. Another
example is the case of spin-orbit interactions of the Rashba type.
These are usually found from the study of the band structure and some
symmetry arguments \cite{Winkler2003}, and they appear generically
as linear terms in the momentum $L_{\text{s.o.}}=-\alpha_{ij}\left(x\right)p_{i}\sigma_{j}$
with some tensor $\alpha_{ij}$ eventually depending in space. In
that case, we convert the momentum operator $p_{j}=-\mathbf{i}\hbar\partial_{j}$
in the space representation, and $L_{\text{s.o.}}$ can be written
as a gauge-potential in the space-sector: $a_{i}^{j}=\alpha_{ij}$
with $g=-m/\hbar$ whereas the Abelian gauge-potential reads $\varphi=\alpha_{ij}^{2}$
with a charge $e=-m/\hbar c$. Obviously, there are freedom in the
choice of the charge and the gauge-potential. Above we gave the natural
notations, when one restores the usual electromagnetism in the Abelian
sector. Contrary to the Abelian situation when a space-time independent
gauge-potential leads to the trivial situation without gauge-field,
the non-Abelian gauge-potential can be space-time independent, yet
the associated gauge-field is non-zero because of the commutator in
the definition \prettyref{eq:F-bold}. For instance, suppose $\alpha_{ij}$
to be space-time independent, then $F_{ij}^{k}=\left(m/\hbar\right)^{2}\alpha_{im}\alpha_{jn}\epsilon^{mnk}$
; note $F_{ij}=0$ when $\alpha_{ij}$ has only one non-zero value
and the one-dimensional spin-orbit problem appears trivial in the
gauge formalism. More complicated spin-orbit interaction, like the
Dresselhaus one which scales as a cubic momentum \cite{Winkler2003}
will not be discussed in our gauge formalism since we started from
the non-relativistic and free quasi-particle model, see \prettyref{eq:H-0}.
In that case, the Fermi surface is isotropic and only linear-in-momentum
spin-orbit interaction -- \textit{i.e.} Rashba-like -- can be described
in a gauge covariant way. The following calculations could nevertheless
be extended to higher order derivatives in principle, see \prettyref{sec:ANNEX-Correspondence-rules}.

When using the Green functions representations \prettyref{eq:electric}
and \prettyref{eq:spin} on their right-hand-side, \prettyref{eq:motion-F-Ab}
and \prettyref{eq:motion-F-NAb} constitute a self-consistent set
of equations of motion for the gauge-potentials, up to the gauge redundancy.
With the Dyson-Gor'kov equations of motion \prettyref{eq:Dyson-1}
and \prettyref{eq:Dyson-2} for the Green-Gor'kov functions, they
constitute a \textit{closed} system of non-linear coupled equations
of motion, which could serve as a basic set of equations for the study
of magnetic superconductivity. Instead of venturing in the perilous
-- and certainly impossible -- task to solve the above system, we
will reduce the complexity of the Dyson-Gor'kov equations in the next
sections. The strategy is to write some quasi-classical expansion
for the Dyson-Gor'kov equations, which will then look like some transport
equations, perhaps easier to solve. At least we will cure the Green-Gor'kov
formalism from its intrinsic difficulty to deal with the evolution
of some two-points correlation functions.

\section{Transport equations\label{sec:Transport-equations}}

We have seen in \prettyref{sec:Green-Gorkov-equation} and \prettyref{sec:Definition-and-Convention}
that the electronic spin degree of freedom can be properly described
in terms of a (non-Abelian) gauge theory. In this formalism, one associates
a curvature -- the gauge field -- with the spin space. To simplify
the Dyson-Gor'kov equations, a usual procedure is to transform the
two-point correlators in the real space to some correlators in the
phase-space, via the Wigner transformation \cite{Hillery1984,Polkovnikov2009}.
Nevertheless, the curvature in the spin space alters the Wigner transformation:
we need a correct transformation of the covariant derivative. This
transformation is pretty lengthy and is given in \prettyref{sec:ANNEX-Correspondence-rules},
in addition to some general recipes for the transformation of the
equations of motion. We here introduce the gauge-covariant Wigner
transformation, and discuss it at the quasi-classical level. Then
we invoke the results of \prettyref{sec:ANNEX-Correspondence-rules},
and we derive a transport-like equation for the quasi-classical propagator
in the phase-space.

The Wigner transformation of the Green function $G\left(x_{1},x_{2}\right)$
(also called the mixed-Fourier transformation) is defined as
\begin{equation}
G\left(p,x\right)=\int dz\left[e^{-\mathbf{i}p\cdot z/\hbar}G\left(x-z/2,x+z/2\right)\right]\label{eq:G-Wigner}
\end{equation}
where $p\cdot z=p_{\mu}z^{\mu}=Et-\boldsymbol{p\cdot x}$ in space-time
(see the beginning of \prettyref{sec:Green-Gorkov-equation}). To
simplify the discussion, we discuss here a generic Green function,
not necessarily the Green-Gor'kov ones introduced in \prettyref{sec:Green-Gorkov-equation}.
We do not introduce a different notation for the Green function $G\left(p,x\right)$
in the phase-space and the correlation function $G\left(x_{1},x_{2}\right)$
in the real space, since the names of their variables are sufficient
to distinguish them. The above definition is obviously not gauge-covariant,
since the Green function transforms as $G\left(x_{1},x_{2}\right)\rightsquigarrow R\left(x_{1}\right)G\left(x_{1},x_{2}\right)R^{\dagger}\left(x_{2}\right)$
and the transformations matrices $R$ are not compensated. We need
a way to get read off the $x_{1,2}$ dependency of the gauge transformation
of the $G\left(x_{1},x_{2}\right)$ function. This is done when one
slightly generalizes \prettyref{eq:G-Wigner} toward a gauge-covariant
Wigner transformation, as we discuss in the next few paragraphs.

The gauge-covariant Wigner transformation has a long and rich history,
and appeared in several places and for different purposes \cite{Stratonovich1956,Fujita1966,Bialynicki-Birula1977,Elze1986,Serimaa1986}.
Most of the studies are devoted to the Abelian gauge theory, when
the gauge-field is supposed classical \cite{Vasak1987,Levanda1994,Levanda2001}
or quantized \cite{Serimaa1986,Javanainen1987}. To the best of my
knowledge, only a few studies are devoted to the non-Abelian problem
of finding a correct gauge-covariant Wigner transformation \cite{Winter1984,Elze1986,Elze1986a,Elze1989},
and none of them address the question of non-relativistic systems.
We here follow the approach of Elze, Gyulassy and Vasak \cite{Elze1986}
who rewrite the Wigner transformation as
\begin{equation}
G\left(p,x\right)=\int dz\left[e^{-\mathbf{i}p\cdot z/\hbar}e^{-z\cdot\partial/2}G\left(x,x\right)e^{z\cdot\partial^{\dagger}/2}\right]\label{eq:G-Wigner-dif}
\end{equation}
where $e^{-z\cdot\partial/2}\Psi\left(x\right)=\Psi\left(x-z/2\right)$
and the same for the derivative $\partial^{\dagger}$ applied to the
left on $\Psi^{\dagger}$, where the fields $\Psi$ are the fermionic
particle-field operators defining the Green function. If one defines
$x_{1,2}=x\mp z/2$, the definitions \prettyref{eq:G-Wigner} and
\prettyref{eq:G-Wigner-dif} are equivalent%
\footnote{In practice, we should include the displacement operators $e^{-z\cdot\partial/2}$
inside the averaging brackets in the definition of the Green functions:
$G\left(p,x\right)=\int dz\left[e^{-\mathbf{i}p\cdot z/\hbar}\left\langle \hat{T}\left[e^{-z\cdot\partial/2}\Psi\left(x\right)\right]\left[e^{z\cdot\partial/2}\Psi\left(x\right)\right]^{\dagger}\right\rangle \right]$
in order to properly define the Wigner transformation of the Green
function. This more rigorous definition nevertheless makes the notations
cumbersome, the reason why we adopt the notations in \prettyref{eq:G-Wigner-dif}.%
}. Additionally, a gauge-covariant Wigner transformation would simply
be deduced from \prettyref{eq:G-Wigner-dif} by the substitution of
the usual derivatives with the covariant ones. Then one defines 
\begin{equation}
G\left(p,x\right)=\int dz\left[e^{-\mathbf{i}p\cdot z/\hbar}e^{-z\cdot D/2}G\left(x,x\right)e^{z\cdot D^{\dagger}/2}\right]\label{eq:G-Wigner-DEF}
\end{equation}
as a gauge-covariant Wigner transformation \cite{Elze1986}, with
a generic covariant derivative $D_{\mu}=\partial_{\mu}+\mathbf{i}A_{\mu}$
for the moment. In a few paragraphs we will come back to the superconductors,
and its bold notations. By definition of $D$ and $D^{\dagger}$,
we have $G\left(p,x\right)\rightsquigarrow R\left(x\right)G\left(p,x\right)R^{\dagger}\left(x\right)$
under a gauge transformation, the property we were looking for. When
the gauge-field is trivial, the definition \prettyref{eq:G-Wigner-DEF}
obviously reduces to \prettyref{eq:G-Wigner-dif}, and so we should
adopt \prettyref{eq:G-Wigner-DEF} as the most general definition
for the gauge-covariant Wigner transformation \cite{Elze1986}. Nevertheless,
it is important to realize that the Wigner transformation \prettyref{eq:G-Wigner-DEF}
has nothing to do with a Fourier transformation anymore, except for
a trivial gauge-field, when \prettyref{eq:G-Wigner-DEF} reduces to
\prettyref{eq:G-Wigner}.

Next one has the property demonstrated in \cite{Elze1986}  
\begin{equation}
e^{-z\cdot D/2}\Psi\left(x\right)=U\left(x,x-z/2\right)\Psi\left(x-z/2\right)\label{eq:U-Psi}
\end{equation}
with 
\begin{equation}
U\left(b,a\right)=\hat{P}\exp\left[-\mathbf{i}\left(b-a\right)^{\mu}\int_{0}^{1}ds\left[A_{\mu}\left(\tau_{s}\right)\right]\right]\label{eq:U-DEF}
\end{equation}
the parallel transport operator \textit{along a straight line} $\tau_{s}=a+\left(b-a\right)s$
parameterized by $s$. The operator $\hat{P}$ orders the path. Injecting
the definition \prettyref{eq:U-DEF}, one rewrites \prettyref{eq:G-Wigner-DEF}
as 
\begin{equation}
G\left(p,x\right)=\int dz\left[e^{-\mathbf{i}p\cdot z/\hbar}U\left(x,x_{1}\right)G\left(x_{1},x_{2}\right)U\left(x_{2},x\right)\right]\label{eq:G-Wigner-UGU}
\end{equation}
in a mixed notation in term of $\left(x,z\right)$ and $\left(x_{1}=x-z/2,x_{2}=x+z/2\right)$
for compactness purpose. The main advantage of promoting \prettyref{eq:G-Wigner-DEF}
instead of \prettyref{eq:G-Wigner-UGU} as the genuine definition
of the gauge-covariant Wigner transformation is because \prettyref{eq:G-Wigner-DEF}
is independent of the path chosen to link the different points $x$
and $x_{1,2}$, whereas there always is an ambiguity in the notation
\prettyref{eq:G-Wigner-UGU}. Because of the definition of the covariant
derivative the paths connecting the points $x_{1}$ to $x$ and from
$x$ to $x_{2}$ are straight lines, as demonstrated and discussed
in \cite{Elze1986,Vasak1987}. Note that the expression \prettyref{eq:G-Wigner-UGU}
was used by Gorini \textit{et al.} \cite{Gorini2010} as a heuristic
definition for a gauge-covariant Wigner transformation, where they
choose a straight line as the simplest realization of the path connecting
the points $x$ to $x_{1}$ or $x_{2}$.

Suppose for a while that $A_{\mu}$ describes a non-trivial Abelian
gauge-field instead of the more elaborated situation of a non-Abelian
problem. Then $A_{\mu}$ commutes with everything, and the definition
\prettyref{eq:U-DEF} reduces to a phase-shift which commutes with
the Green correlation function in \prettyref{eq:G-Wigner-UGU}. In
that case the two phase shifts $U\left(x,x_{1}\right)$ and $U\left(x_{2},x\right)$
combine in a resulting Abelian phase shift 
\begin{equation}
U_{\text{Abel.}}\left(x_{1},x_{2}\right)=e^{-\mathbf{i}z^{\mu}\int_{0}^{1}ds\left[A_{\mu}\left(x+z\left(s-1\right)/2\right)\right]}
\end{equation}
where the path connects now the points $x_{1}$ and $x_{2}$. This
Abelian phase shift has been used in the description of a gauge-invariant
Wigner transformation when only electromagnetism is taken into account
\cite{Serimaa1986,Javanainen1987,Vasak1987,Levanda1994,Levanda2001}.
The above Abelian phase shift is also sometimes heuristically introduced
in order to obtain some gauge-covariant Wigner transformation for
both the normal metal \cite{Fujita1966} and the superconductor \cite{b.kopnin}
situations. Note that $U_{\text{Abel.}}\left(x_{1},x_{2}\right)$
is gauge \textit{in}variant, whereas the Wigner transformation \prettyref{eq:G-Wigner-UGU}
is gauge \textit{co}variant. This is the main difference between Abelian
and non-Abelian gauge theory: in the later case, there is no gauge
invariant quantity in the theory, only the observables which trace
out the gauge degrees of freedom are gauge invariant, see also below
the construction of the current densities in \prettyref{eq:neutral-transport},
\prettyref{eq:electric-transport} and \prettyref{eq:spin-transport},
and more general literature on this subject \cite{Peskin1995}. 

Equipped with the above gauge-covariant Wigner transformation \prettyref{eq:G-Wigner-UGU},
we can now come back to the problem of the obtention of the transport
equations for a non-Abelian superconducting plasma. The strategy is
to start with the Dyson equations of motion \prettyref{eq:Dyson-1}
and \prettyref{eq:Dyson-2} and the propagator \prettyref{eq:G0-covariant}.
Then we transform the two Dyson equations according to the Wigner
transformation 
\begin{equation}
\mathbf{G}\left(p,x\right)=\int dz\left[e^{-\mathbf{i}p\cdot z/\hbar}\mathbf{U}\left(x,x_{1}\right)\mathbf{G}\left(x_{1},x_{2}\right)\mathbf{U}\left(x_{2},x\right)\right]\label{eq:G-Wigner-UGU-bold}
\end{equation}
with
\begin{equation}
\mathbf{U}\left(b,a\right)=\hat{P}\exp\left[-\mathbf{i}\left(b-a\right)^{\mu}\int_{0}^{1}ds\left[\mathbf{A}_{\mu}\left(\tau_{s}\right)\right]\right]
\end{equation}
the parralel transport in the $U\left(1\right)\otimes SU\left(2\right)\otimes SU\left(2\right)$
space, when the gauge-potential $\mathbf{A}_{\mu}$ is defined in
\prettyref{eq:A-def}. The propagator \prettyref{eq:G0-covariant}
contains the spin texture in a covariant manner, and the associated
curvature is properly taken into account in the definition \prettyref{eq:G-Wigner-UGU-bold}.
The covariant derivatives in $\mathbf{G}^{-1}\left(x\right)$ can
be transformed according to the set of rules found in \prettyref{sec:ANNEX-Correspondence-rules}.
Since the gauge-potential $\mathbf{A}_{\mu}$ is diagonal in the Nambu
space, it commutes with $\tau_{3}$, as well as the gauge-field \prettyref{eq:F-bold}.
One then immediately has that
\begin{multline}
\int dz\left[e^{-\mathbf{i}p\cdot z/\hbar}\mathbf{U}\left(\tau_{3}\mathbf{D}_{0}\mathbf{G}\left(x_{1},x_{2}\right)\right)\mathbf{U}\right]=\\
\tau_{3}\int dz\left[e^{-\mathbf{i}p\cdot z/\hbar}\mathbf{U}\left(\mathbf{D}_{0}\mathbf{G}\left(x_{1},x_{2}\right)\right)\mathbf{U}\right]
\end{multline}
which greatly simplifies the following treatment. The parallel transport
operators $\mathbf{U}\left(x,x_{1}\right)$ on the left and $\mathbf{U}\left(x_{2},x\right)$
on the right always have the same space-time dependencies, so we do
not write them explicitly. Finally, the pair potential is treated
as a conventional potential in the $SU\left(2\right)\otimes SU\left(2\right)\otimes U\left(1\right)$
space, according to the general recipe
\begin{multline}
\int dz\left[e^{-\mathbf{i}p\cdot z/\hbar}\mathbf{U}\left(\mathbf{M}\left(x_{1}\right)\mathbf{G}\left(x_{1},x_{2}\right)\right)\mathbf{U}\right]=\\
\mathbf{U}\left(x,x-\mathbf{i}\hbar\partial_{p}/2\right)\mathbf{M}\left(x-\mathbf{i}\hbar\partial_{p}/2\right)\mathbf{U}\left(x-\mathbf{i}\hbar\partial_{p}/2,x\right)\mathbf{G}\\
\approx\mathbf{M}\left(x\right)\mathbf{G}\left(p,x\right)-\mathbf{i}\dfrac{\hslash}{2}\boldsymbol{\mathfrak{D}}_{\nu}\mathbf{M}\left(x\right)\partial_{p}^{\nu}\mathbf{G}\left(p,x\right)\label{eq:G-Wigner-UMGU}
\end{multline}
at first order in $\hbar$, where we defined the covariant derivative
\begin{equation}
\boldsymbol{\mathfrak{D}}_{\nu}\mathbf{M}=\partial_{\nu}\mathbf{M}+\mathbf{i}\left[\mathbf{A}_{\nu},\mathbf{M}\right]\label{eq:cov-deriv-super-2tensor}
\end{equation}
applied to any matrix $\mathbf{M}\in SU\left(2\right)\otimes SU\left(2\right)\otimes U\left(1\right)$.
We injected some $\mathbf{U}\left(x_{1},x\right)\mathbf{U}\left(x,x_{1}\right)=1$
in the Wigner transformation \prettyref{eq:G-Wigner-UMGU}, thanks
to the straight path convention in \prettyref{eq:U-DEF}. A similar
calculation gives
\begin{multline}
\int dz\left[e^{-\mathbf{i}p\cdot z/\hbar}\mathbf{U}\left(\mathbf{G}\left(x_{1},x_{2}\right)\mathbf{M}\left(x_{2}\right)\right)\mathbf{U}\right]=\\
\approx\mathbf{G}\left(p,x\right)\mathbf{M}\left(x\right)+\mathbf{i}\dfrac{\hslash}{2}\partial_{p}^{\nu}\mathbf{G}\left(p,x\right)\boldsymbol{\mathfrak{D}}_{\nu}\mathbf{M}\left(x\right)
\end{multline}
when the potential is applied on the second variable from the right. 

Taking the difference and the sum of the Dyson equations \prettyref{eq:Dyson-1}
and \prettyref{eq:Dyson-2}, we finally have:
\begin{multline}
\dfrac{\mathbf{i}\hslash c}{2}\left[\tau_{3},\boldsymbol{\mathfrak{D}}_{0}\mathbf{G}\right]_{+}+\hbar\omega\left[\tau_{3},\mathbf{G}\right]_{-}+\mathbf{i}\hslash v^{i}\boldsymbol{\mathfrak{D}}_{i}\mathbf{G}\left(p,x\right)\\
+\left[\boldsymbol{\Delta}\left(x\right),\mathbf{G}\right]_{-}+\mathbf{i}\dfrac{\hslash}{2}\left[\boldsymbol{\mathfrak{D}}_{\mu}\boldsymbol{\Delta},\partial_{p}^{\mu}\mathbf{G}\right]_{+}\\
+\dfrac{\mathbf{i}\hbar}{8}\left[\mathbf{F}_{i0}\left(3\tau_{3}\partial_{p}^{i}\mathbf{G}+\partial_{p}^{i}\mathbf{G}\tau_{3}\right)+\left(\tau_{3}\partial_{p}^{i}\mathbf{G}+3\partial_{p}^{i}\mathbf{G}\tau_{3}\right)\mathbf{F}_{i0}\right]\\
+\mathbf{i}\dfrac{\hslash}{2}v^{i}\left[\mathbf{F}_{\mu i}\left(x\right),\partial_{p}^{\mu}\mathbf{G}\right]_{+}-\left(\mathbf{I}_{+}-\mathbf{I}_{-}\right)=0\label{eq:transport-diff}
\end{multline}
\begin{multline}
\dfrac{\mathbf{i}\hslash c}{2}\left[\tau_{3},\boldsymbol{\mathfrak{D}}_{0}\mathbf{G}\right]_{-}+\hbar\omega\left[\tau_{3},\mathbf{G}\right]_{+}-\dfrac{p^{2}}{m}\mathbf{G}\left(p,x\right)\\
+\left[\boldsymbol{\Delta}\left(x\right),\mathbf{G}\right]_{+}-\mathbf{i}\dfrac{\hslash}{2}\left[\boldsymbol{\mathfrak{D}}_{\mu}\boldsymbol{\Delta},\partial_{p}^{\mu}\mathbf{G}\right]_{-}\\
+\dfrac{\mathbf{i}\hbar}{8}\left[\mathbf{F}_{i0}\left(3\tau_{3}\partial_{p}^{i}\mathbf{G}-\partial_{p}^{i}\mathbf{G}\tau_{3}\right)+\left(\tau_{3}\partial_{p}^{i}\mathbf{G}-3\partial_{p}^{i}\mathbf{G}\tau_{3}\right)\mathbf{F}_{i0}\right]\\
+\mathbf{i}\dfrac{\hslash}{4}v^{i}\left[\mathbf{F}_{\mu i}\left(x\right),\partial_{p}^{\mu}\mathbf{G}\right]_{-}-\left(\mathbf{I}_{+}+\mathbf{I}_{-}\right)=2\label{eq:transport-sum}
\end{multline}
at first order in $\hbar$. We defined $v^{i}=p^{i}/m$ a velocity,
and $\left[A,B\right]_{\pm}=AB\pm BA$ define the (anti-)commutator.
The above transport-like equation \prettyref{eq:transport-diff} is
the quasi-classical equation for superconductors in the presence of
non-Abelian gauge-fields. The sum-equation \prettyref{eq:transport-sum}
helps when discussing the quasi-classical correlation function $\mathbf{G}\left(p,x\right)$
and its quantum corrections. The terms
\begin{multline}
\mathbf{I}_{+}\left(p,x\right)=\int dz\int dye^{-\mathbf{i}p\cdot z/\hbar}\times\\
\left[\mathbf{U}\left(x,x_{1}\right)\boldsymbol{\Sigma}\left(x_{1},y\right)\mathbf{G}\left(y,x_{2}\right)\mathbf{U}\left(x_{2},x\right)\right]
\end{multline}
\begin{multline}
\mathbf{I}_{-}\left(p,x\right)=\int dz\int dye^{-\mathbf{i}p\cdot z/\hbar}\times\\
\left[\mathbf{U}\left(x,x_{1}\right)\mathbf{G}\left(x_{1},y\right)\boldsymbol{\Sigma}^{\dagger}\left(y,x_{2}\right)\mathbf{U}\left(x_{2},x\right)\right]\label{eq:collision-integrals}
\end{multline}
correspond to the impurities scattering terms. They are generically
called collision integrals. Here, they are gauge-covariant by construction.

The difference \prettyref{eq:transport-diff} and sum \prettyref{eq:transport-sum}
equations are obviously covariant with respect to the gauge-transformation
\begin{align}
\mathbf{G}\left(p,x\right) & \rightsquigarrow\mathbf{R}\left(x\right)\mathbf{G}\left(p,x\right)\mathbf{R}^{\dagger}\left(x\right)\nonumber \\
\mathbf{A}_{\mu}\left(x\right) & \rightsquigarrow\mathbf{R}\left(x\right)\mathbf{A}_{\mu}\left(x\right)\mathbf{R}^{\dagger}\left(x\right)-\mathbf{i}\mathbf{R}\left(x\right)\partial_{\mu}\mathbf{R}^{\dagger}\left(x\right)\nonumber \\
\boldsymbol{\Delta}_{\mu\nu}\left(x\right) & \rightsquigarrow\mathbf{R}\left(x\right)\boldsymbol{\Delta}_{\mu\nu}\left(x\right)\mathbf{R}^{\dagger}\left(x\right)\label{eq:gauge-transform-quasiclass}
\end{align}
since $\mathbf{R}$ and $\tau_{3}$ commute. Note that the gauge-transformation
for the quasi-classical Green function is local, in comparison with
\prettyref{eq:gauge-transform-Gxx}. This is due to the definition
of the gauge-covariant Wigner transformation \prettyref{eq:G-Wigner-UGU}.

The Abelian version of these equations reduces to the usual ones \cite{Eckern1981,Aronov1981,serene_rainer.1983,b.kopnin,Kita2001}.
Since $\mathbf{A}_{\mu}$ is diagonal and $A_{\mu}$ is real in the
Abelian case, the Abelian limit corresponds to $\mathbf{F}_{\mu\nu}=\tau_{3}F_{\mu\nu}$
and $F_{\mu\nu}=\partial_{\mu}A_{\nu}-\partial_{\nu}A_{\mu}$ commutes
with $\mathbf{G}$. Note in this case that the covariant derivative
$\boldsymbol{\mathfrak{D}}_{\mu}$ still contains a non-trivial gauge
potential part, responsible for the asymmetry between the $F\left(p,x\right)$
and the $G\left(p,x\right)$ sectors: the space derivative reads $\left(\boldsymbol{\partial_{x}}\pm2\mathbf{i}e\boldsymbol{A}/\hbar\right)$
in front of the anomalous correlation sector, whereas the $G\left(p,x\right)$
correlation function becomes uncharged. Supposing further the absence
of gauge field reduces the above equations to the usual transport
equations for superconductors, when all the gauge fields vanish, and
the covariant derivatives reduce to the usual derivative \cite{Langenberg1986,b.kopnin}. 

The normal metal limit consists in projecting \prettyref{eq:transport-diff}
and \prettyref{eq:transport-sum} to the particle sector, \textit{i.e.}
choosing $\tau_{3}=1$ and $\boldsymbol{\Delta}=0$. We then recover
the non-Abelian case \cite{Gorini2010}. Supposing a pure Abelian
gauge field reduces further to the usual transport equation \cite{Altshuler1979,Langreth1966,Javanainen1987,Serimaa1986,Levanda1994,Levanda2001}.
Interestingly enough, the covariant derivatives are reduced to the
usual derivatives in this case. Finally, the standard transport equations
are recovered when we suppose in addition that the gauge fields disappear
\cite{Kadanoff1962,Rammer1986}.

The equations \prettyref{eq:transport-diff} and \prettyref{eq:transport-sum}
constitute the fundamental result of this study. They are strictly
equivalent to the Dyson-Gor'kov equations \prettyref{eq:Dyson-1}
and \prettyref{eq:Dyson-2} in the low energy sector, characterized
by the relation 
\begin{equation}
\dfrac{\hbar}{\tilde{p}\tilde{x}}\ll1\label{eq:quasiclassic-DEF}
\end{equation}
where $\tilde{p}$ and $\tilde{x}$ stand for the characteristic values
of the momentum and the position, respectively. Additionally, the
variations of the momentum and/or the position must rely on the approximation
\prettyref{eq:quasiclassic-DEF} which then defines the quasi-classic
evolution.

We still have to define the observables associated with the transport
equations. The microscopic quantities \prettyref{eq:neutral}, \prettyref{eq:electric}
and \prettyref{eq:spin} are all evaluated in the limit $x_{1}\rightarrow x_{2}$.
Since it corresponds to the limit $z\rightarrow0$, the Wigner transformation
\prettyref{eq:G-Wigner-UGU} is not well defined, and we have to find
a work-around to obtain the expressions for the charges and current
densities. According to the general recipe in the Abelian case \cite{Kadanoff1962},
we can suppose the density to be the integrated version of the phase-space
density $\mathbf{G}\left(p,x\right)$ over the momentum. Then we propose
to define
\begin{align}
\rho_{n}\left(x\right) & =\hbar\int\dfrac{dp}{2\pi\hbar}\text{Tr}\left\{ \tau_{3}\mathbf{G}\left(p,x\right)\right\} \nonumber \\
\rho_{e}\left(x\right) & =e\hbar\int\dfrac{dp}{2\pi\hbar}\text{Tr}\left\{ \mathbf{G}\left(p,x\right)\right\} \nonumber \\
\rho_{e}\left(x\right) & =\dfrac{g}{2}\hbar\int\dfrac{dp}{2\pi\hbar}\text{Tr}\left\{ \mathbf{s}_{k}\tau_{3}\mathbf{G}\left(p,x\right)\right\} \label{eq:rho-transport}
\end{align}
for the particle, electric and spin densities, respectively, where
we used the notation $dp/2\pi\hbar\equiv dp_{x}dp_{y}dp_{z}d\omega/\left(2\pi\right)^{4}\hbar^{3}c$,
such that the proposed densities $\rho_{n,e,s}$ depend only on space-time.
Our strategy is to manipulate the transport equation \prettyref{eq:transport-diff}
in order to obtain the conservation laws $\partial_{t}\rho_{n,e,s}+\boldsymbol{\partial_{x}\cdot j}_{n,e,s}=0$,
then we identify the conserved current as the correct particle, electric
and spin currents, respectively. Before turning to this program, we
define
\begin{equation}
\Delta_{0}\left(x\right)=-\hbar V_{0}\left(x\right)\int\dfrac{dp}{2\pi\hbar}\text{Tr}\left\{ \mathbf{i}\sigma_{2}\tau_{+}\mathbf{G}\left(p,x\right)\right\} \label{eq:self-consistent-transport}
\end{equation}
for the self-consistent relation (compare with \prettyref{eq:self-consistent}).
Next we take the trace of \prettyref{eq:transport-diff}, which cancels
most of the terms. Then we integrate over the momentum, which cancels
all the gauge field terms and the collision integrals. We are left
with the integrated trace of $\left(c\boldsymbol{\mathfrak{D}}_{0}\tau_{3}+v^{i}\boldsymbol{\mathfrak{D}}_{i}\right)\mathbf{G}\left(p,x\right)$.
The commutators in the covariant derivatives cancel thanks to the
cyclic properties of the trace, and we obtain the desired continuity
equation, with the conserved current 
\begin{equation}
j_{n}^{i}=\hbar\int\dfrac{dp}{2\pi\hbar}\text{Tr}\left\{ v^{i}\mathbf{G}\left(p,x\right)\right\} \label{eq:neutral-transport}
\end{equation}
for the quasiparticles current density. A similar calculation after
replacement of $\mathbf{G}\left(p,x\right)$ by $\tau_{3}\mathbf{G}\left(p,x\right)$
in \prettyref{eq:transport-diff} leads to the conserved current
\begin{equation}
j_{e}^{i}=e\hbar\int\dfrac{dp}{2\pi\hbar}\text{Tr}\left\{ \tau_{3}v^{i}\mathbf{G}\left(p,x\right)\right\} \label{eq:electric-transport}
\end{equation}
for the electric current density. To obtain the spin current, we replace
$\mathbf{G}\left(p,x\right)$ by $\mathbf{s}_{k}\mathbf{G}\left(p,x\right)$
in \prettyref{eq:transport-diff}, we take the trace and we integrate
over the momentum. Nevertheless, the gap terms remain. They cancel
in virtue of the relation \prettyref{eq:self-consistent-transport}
and the cyclic property of the trace. We are left with
\begin{equation}
j_{s}^{ik}=\hbar\dfrac{g}{2}\int\dfrac{dp}{2\pi\hbar}\text{Tr}\left\{ \mathbf{s}_{k}v^{i}\mathbf{G}\left(p,x\right)\right\} \label{eq:spin-transport}
\end{equation}
for the spin current density. Relations \prettyref{eq:rho-transport},
\prettyref{eq:neutral-transport}, \prettyref{eq:electric-transport}
and \prettyref{eq:spin-transport} are the observables associated
with the equations of motion \prettyref{eq:transport-diff} and \prettyref{eq:transport-sum}.
Injected in the relations \prettyref{eq:motion-F-Ab} and \prettyref{eq:motion-F-NAb}
they constitute a complete set of self-consistent equations for the
superconducting plasma, provided we use the self-consistency relation
\prettyref{eq:self-consistent-transport} and the definitions \prettyref{eq:F-charge-def}
and \prettyref{eq:F-spin-def} in addition to the gauge redundancy
\prettyref{eq:gauge-transform-quasiclass}. In comparison with the
normal metal when only electromagnetism is present, one realizes that
the gauge potential appears alongside the gauge field in the transport
equation for the superconductors, a clear hallmark of their quantum
behavior. The non-Abelian generalization we provided here even enriches
this picture, with possible magneto-electric couplings and non-trivial
boundary conditions at the interface between different devices.

Let us now discuss the inclusion of impurities into the transport
equation in term of the self-energy via the substitution \prettyref{eq:G-1-self-energy}.
The explicit form of the self-energy depends on the materials one
wants to describe, and on the approximation one develops for it. For
a disorder weakly coupled to the particle trajectories and randomly
distributed along the sample, the Born approximation might be sufficient.
When the disorder is also isotropic, the self-energy term becomes
constant in momentum \cite{b.abrikosov_gorkov,b.kopnin}
\begin{equation}
\boldsymbol{\Sigma}\left(\omega,x\right)=\dfrac{\hbar}{2\pi N_{0}^{d}\tau c}\int\dfrac{d^{d}\boldsymbol{p}}{\left(2\pi\hbar\right)^{d}}\mathbf{G}\left(p,x\right)\label{eq:impurities-Born-isotropic}
\end{equation}
with $N_{0}$ the density of particles in the normal state which depends
on the space dimension $d$, and $\tau$ the mean free time. We will
not explore the effect of the disorder beyond this simple model in
the following.

To conclude this section, we remark that an alternative way toward
the quasi-classical superconducting equations is to construct some
propagator
\begin{equation}
\mathbf{G}^{-1}\left(x\right)=\mathbf{i}\hslash\tau_{3}\boldsymbol{\Pi}_{0}\left(x\right)-\dfrac{\hslash^{2}}{2m}\mathbf{D}_{j}\mathbf{D}^{j}\left(x\right)
\end{equation}
instead of \prettyref{eq:G0-covariant}, with an alternative covariant
derivative in the time sector 
\begin{align}
\boldsymbol{\Pi}_{0} & =\partial_{0}+\mathbf{i}\tau_{3}\left(\begin{array}{cc}
A_{0} & \Delta\\
\Delta^{\dagger} & A_{0}^{\ast}
\end{array}\right)=\partial_{0}+\mathbf{i}\mathfrak{B}_{0}\left(x\right)
\end{align}
where the pair potential has been promoted to be a gauge potential
in the time sector of the Nambu space. Working with the $\boldsymbol{\Pi}_{0}$
operator, one has to define a parallel displacement as
\begin{equation}
\tilde{\mathbf{U}}\left(b,a\right)=\hat{P}\exp\left[-\mathbf{i}\int_{a}^{b}dz^{\mu}\mathfrak{B}_{\mu}\left(z\right)\right]\label{eq:U-B-alternative}
\end{equation}
with $\mathfrak{B}_{\mu}\equiv\left(\mathfrak{B}_{0},-\mathbf{A}_{j}\right)$,
and we have to use an expression like 
\begin{multline}
\int dz\left[e^{-\mathbf{i}p\cdot z/\hbar}\tilde{\mathbf{U}}\left(\tau_{3}\boldsymbol{\Pi}_{0}\left(x_{1}\right)\mathbf{G}\right)\tilde{\mathbf{U}}\right]=\\
\tilde{\mathbf{U}}\left(x,x-\mathbf{i}\hbar\partial_{p}\right)\tau_{3}\tilde{\mathbf{U}}\left(x-\mathbf{i}\hbar\partial_{p},x\right)\\
\times\int dz\left[e^{-\mathbf{i}p\cdot z/\hbar}\tilde{\mathbf{U}}\left(\boldsymbol{\Pi}_{0}\left(x_{1}\right)\mathbf{G}\left(x_{1},x_{2}\right)\right)\tilde{\mathbf{U}}\right]
\end{multline}
which is more complicated to deal with than the convention \prettyref{eq:G-Wigner-UGU-bold}
we used before, since the operator $\tau_{3}$ does not commute with
the parallel displacement operator $\tilde{\mathbf{U}}$ in \prettyref{eq:U-B-alternative}
anymore. Even in the quasi-classical limit, the resulting transport
equations will not look like \prettyref{eq:transport-diff} and \prettyref{eq:transport-sum},
since now the gap parameter has the property of a gauge potential,
and accordingly transforms like $\mathfrak{B}_{0}\rightsquigarrow\mathbf{R}\mathfrak{B}_{0}\mathbf{R}^{\dagger}-\mathbf{i}\mathbf{R}\partial_{ct}\mathbf{R}^{\dagger}$
under a gauge transformation. This property may have interesting consequences
-- especially for the symmetry classification of superconducting states
for instance -- that we keep for future studies. Note also that a
further generalization of the gap parameter could make possible its
inclusion as some gauge potential in the space sector (some terms
in the $\boldsymbol{\Pi}_{i}=\partial_{i}-\mathbf{i}\mathfrak{B}_{i}\left(x\right)$
which are absent in our present construction), which seems to take
into account higher symmetries of the gap parameter ($p$-wave for
instance), see \textit{e.g.} \cite{b.mineev_samokhin} for the usual
treatment of such symmetries. This hypothesis is far beyond the scope
of this introductory study.

\section{Eilenberger equation\label{sec:Eilenberger-equation}}

In this section, we simplify even more the equation of motion for
the quasi-classical Green functions, towards the so-called Eilenberger
equation, here generalized to include non-Abelian gauge interactions.
Reader familiar with the usual derivation of the quasi-classical equation
\cite{eilenberger.1968,larkin_ovchinnikov.1969,serene_rainer.1983,b.kopnin,Gorkov2008}
for superconductors can just have a look on the expression \prettyref{eq:Eilenberger}
and skip the remaining of this section.

The transport equation \prettyref{eq:transport-diff} was valid at
first order in $\hbar/\tilde{p}\tilde{x}$, where $\tilde{p}$ and
$\tilde{x}$ are characteristic values for the momentum and the space
variations. The characteristic values for a superconductor are the
Fermi momentum $p_{F}$ and the coherence length $\xi_{0}$, verifying
\begin{equation}
\dfrac{\hbar}{p_{F}\xi_{0}}\sim\dfrac{\lambda_{F}}{\xi_{0}}\sim\dfrac{\Delta}{E_{F}}\ll1\label{eq:quasiclassic-supra}
\end{equation}
in most of the cases. This means that, for a description in space
with resolution $\xi_{0}$ at best, one can content ourself with fixing
the momentum to be the Fermi one in \prettyref{eq:transport-diff}.
Thus, the amplitude of the momentum is pinned to the Fermi surface,
and we could forget all the momentum derivatives in the transport
equations \prettyref{eq:transport-diff} and \prettyref{eq:transport-sum}.
Nevertheless, the angular dependency of the momentum is still free
in principle. For instance, suppose a two-dimensional and circular
Fermi surface (a Fermi circle then), we decompose $\boldsymbol{p}=p_{F}\boldsymbol{\hat{p}_{F}}+p_{\varphi}\boldsymbol{\hat{\varphi}}$,
with unit radial $\boldsymbol{\hat{p}_{F}}$ and tangential $\boldsymbol{\hat{\varphi}}$
vectors. Next, the gradient in the momentum space reads $\boldsymbol{\partial_{p}}=\boldsymbol{\hat{p}_{F}}\boldsymbol{\partial_{\hat{p}}}+p_{F}^{-1}\boldsymbol{\hat{\varphi}}\boldsymbol{\partial_{\varphi}}$
and we suppose that the variation along the radial direction $\boldsymbol{\hat{p}_{F}}$
vanishes. Then we note that the contribution $\hbar/p_{F}\ll\xi_{0}$
is small, and the radial derivative should be neglected as well. This
situation is generic, and valid for three-dimensional problems as
well as for non symmetrical Fermi surfaces, see \cite{b.kopnin} for
longer discussions.

So all the momentum derivatives \prettyref{eq:transport-diff} are
of order of magnitude at least $\hbar p_{F}^{-1}$ or even higher,
and we are left with 
\begin{multline}
\dfrac{\mathbf{i}\hbar c}{2}\left[\tau_{3},\boldsymbol{\mathfrak{D}}_{0}\mathbf{G}\right]_{+}+\hbar\omega\left[\tau_{3},\mathbf{G}\right]_{-}+\mathbf{i}\hbar v_{F}^{i}\boldsymbol{\mathfrak{D}}_{i}\mathbf{G}\\
+\left[\boldsymbol{\Delta}\left(x\right),\mathbf{G}\right]_{-}=\mathbf{I}_{+}-\mathbf{I}_{-}\label{eq:Eilenberger-G}
\end{multline}
for the transport equation with relevant energies at the Fermi level.
The phase-space dependency now is confined to the Fermi surface in
momentum, whereas the frequency is well below the superconducting
gap. The gauge-potentials should be of low energy, so their characteristic
length should be larger than the coherence one, too. The collision
integrals will be discussed later.

Usually one replaces $p/m\approx v_{F}$, the Fermi velocity, in front
of the space derivative, as we naively did in \prettyref{eq:Eilenberger-G}.
Here one might wonder whether the Fermi surface is a well defined
quantity. Indeed, it is well known that adding a spin-orbit and/or
a Zeeman effect splits the Fermi surface in two sheets \cite{Winkler2003}.
Nevertheless, the gauge theory is an extension of the free particles
model (see \prettyref{eq:G0-covariant}), for which the Fermi surface
has a single sheet. So the genuine Fermi surface in \prettyref{eq:Eilenberger-G}
is the conventional one defined for the free-particles. Despite the
absence of a Fermi surface \textit{per se} for superconductors, one
can still replace $p/m\approx v_{F}$ with $v_{F}$ the Fermi velocity
for the free particles defined in the absence of the Cooper pairing.
The only momentum dependency which remains is the angular one.

Still because the gap parameter is weak in comparison with the Fermi
energy, all the characteristic energies in \prettyref{eq:Eilenberger-G}
will be close to the Fermi energy. Then we can define a renormalized
quasi-classical function, the so-called $\xi$-integrated Green function
\begin{equation}
\mathbf{g}\left(\Omega_{F},\omega,x\right)=\int\dfrac{d\xi_{p}}{\mathbf{i}\pi c}\mathbf{G}\left(p,x\right)
\end{equation}
with $d\xi_{p}=v_{F}dp$, the increment of the linear variation of
the energy relative to the Fermi one: $\xi_{p}=v_{F}\left(p-p_{F}\right)$,
see \cite{b.abrikosov_gorkov,b.kopnin} for more details about the
quantity $\xi_{p}$. Note that $\mathbf{g}$ still depends on the
solid angle $\Omega_{F}$ in the momentum space at the Fermi surface.
In a sense, $\mathbf{g}$ corresponds to the low-energy sector of
the quasi-classical Green function, when the high-energies have been
integrated out. It sometimes requires some care to explicitely make
this integration, see \textit{e.g.} \cite{serene_rainer.1983,b.kopnin}.
Once $\xi$-integrated, the equation \prettyref{eq:Eilenberger-G}
is called the Eilenberger equation \cite{eilenberger.1968,larkin_ovchinnikov.1969,b.kopnin}.
The transposition of \prettyref{eq:Eilenberger-G} toward the $\xi$-integrated
representation of the quasi-classical Green function is straightforward,
except for the collision integral which we discuss now separately.

If we suppose the disorder to be weakly and isotropically interacting
with the electrons and randomly distributed along the sample, a convenient
approximation to describe it is the Born approximation \prettyref{eq:impurities-Born-isotropic}.
There, we substitute $d^{3}p/\left(2\pi\hbar\right)^{3}\approx N_{0}^{3d}d\xi_{p}d\Omega_{p}/4\pi$
in 3D, $d^{2}p/\left(2\pi\hbar\right)^{2}\approx N_{0}^{2d}d\xi_{p}d\Omega_{p}/2\pi$
in 2D or $dp/2\pi\hbar=N_{0}^{1d}d\xi_{p}$ in 1D, with $\Omega_{p}$
the solid-angle in the momentum space and $N_{0}^{3d}=mp_{F}/2\pi^{2}\hbar^{3}$,
$N_{0}^{2d}=m/2\pi\hbar^{2}$ or $N_{0}^{1d}=m/p_{F}2\pi\hbar$ the
density of state in the normal metal in 3D, 2D and 1D, respectively.
In the following we treat the 3D case and note it $N_{0}$ since the
substitution are straightforward for the other dimensions. The integral
in \prettyref{eq:impurities-Born-isotropic} then reduces to
\begin{equation}
\boldsymbol{\Sigma}\left(x,\omega\right)=\dfrac{\mathbf{i}\hbar}{2\tau}\left\langle \mathbf{g}\right\rangle \label{eq:self-energy-Eilenberger}
\end{equation}
where $\left\langle \cdots\right\rangle $ stands for the averaging
of the quasi-classical $\xi$-integrated functions over the Fermi
surface, spanned by the increment $d\Omega_{p}$:
\begin{equation}
\left\langle \mathbf{g}\right\rangle =\int\dfrac{d\Omega_{p}}{4\pi}\mathbf{g}\left(\Omega_{F},\omega,x\right)\label{eq:averaging-Fermi-surf}
\end{equation}
and so on for 2D and 1D, where the average is just the sum over two
contributions.

We can then integrate the equation \prettyref{eq:Eilenberger-G} over
the energies $d\xi_{p}$. Since the self-energy is already $\xi$-integrated
by virtue of the relation \prettyref{eq:self-energy-Eilenberger}
and more generally by the use of the Born approximation, the integration
consists in the replacement of the $\mathbf{G}\left(p,x\right)$ function
by the $\mathbf{g}\left(\Omega_{F},\omega,x\right)$ one. Then the
Eilenberger equation reads
\begin{multline}
\dfrac{\mathbf{i}\hbar c}{2}\left[\tau_{3},\boldsymbol{\mathfrak{D}}_{0}\mathbf{g}\right]_{+}+\mathbf{i}\hbar v^{i}\boldsymbol{\mathfrak{D}}_{i}\mathbf{g}\left(\Omega_{F},\omega,x\right)\\
+\left[\hbar\omega\tau_{3}+\boldsymbol{\Delta}-\dfrac{\mathbf{i}\hbar}{2\tau}\left\langle \mathbf{g}\right\rangle ,\mathbf{g}\right]_{-}=0\label{eq:Eilenberger}
\end{multline}
in the non-Abelian case. In the case of a simpler Abelian gauge field,
one has $\mathbf{A}_{\mu}=e\tau_{3}A_{\mu}/\hbar$ with $A_{\mu}$
real, and the equation looks exactly the same \cite{b.kopnin,Tanaka2009b}.
The commutator in the covariant derivative distributes the charge
asymmetrically among the components of $\mathbf{g}$, and the equation
for $g$ looks uncharged in the Abelian case. In the absence of a
gauge field, the covariant derivatives disappear, and only normal
derivatives remain, see \textit{e.g.} \cite{b.kopnin} for these two
situations. 

The generalized Eilenberger equation \prettyref{eq:Eilenberger} consistent
with a non-Abelian gauge theory is our second important result in
this study. It may allow considerable simplifications in the understanding
of intricate problems dealing with spin textures and superconductivity.

Being a homogeneous equation, the Eilenberger equation \prettyref{eq:Eilenberger}
accepts all multiples of $\mathbf{g}$ as solution. In addition, the
restriction $\hbar/p_{F}\xi_{0}\ll1$ makes the sum-equation \prettyref{eq:transport-sum}
meaningless, as can be checked easily after cancelation of all the
terms we discarded in this section: it gives the classical expression
for the $\xi$-integrated function. The remedy to this curse is the
so-called normalization condition \cite{eilenberger.1968,b.kopnin}.
Multiplying \prettyref{eq:Eilenberger} from the left by $\mathbf{g}$,
or from the right by $\mathbf{g}$, and then summing the two contributions,
we realize that the commutator helps making both $\mathbf{g}$ and
$\mathbf{gg}$ solutions of the same equation \prettyref{eq:Eilenberger}.
This means that a generic solution reads $\mathbf{gg}=A\mathbf{g}+B$,
with $A$ and $B$ two constants \cite{b.kopnin}. The normalization
condition for $\mathbf{g}\left(\Omega_{F},\omega,x\right)$ reads
then
\begin{equation}
\mathbf{gg}=1\label{eq:normalisation-condition}
\end{equation}
found as the solution of \prettyref{eq:Eilenberger} for large time
and space, where there are neither impurities nor gauge-field, and
when the $\xi$-integration can be performed exactly, for which situation
we find $A=1$ and $B=0$ \cite{b.kopnin}.

We shortly give the definitions of the observables for the $\xi$-integrated
functions. They follow from the substitution of the integration element
$dp/2\pi\hbar\approx N_{0}d\xi_{p}\left(d\Omega_{p}/4\pi\right)\left(d\omega/2\pi\hbar c\right)$
in the general relations \prettyref{eq:rho-transport}, \prettyref{eq:neutral-transport},
\prettyref{eq:electric-transport} and \prettyref{eq:spin-transport}.
One has
\begin{align}
\rho_{n}\left(x\right) & =\mathbf{i}\pi\hbar N_{0}\int\dfrac{d\omega}{2\pi}\left\langle \text{Tr}\left\{ \tau_{3}\mathbf{g}\right\} \right\rangle \nonumber \\
j_{n}^{i}\left(x\right) & =\mathbf{i}\pi\hbar N_{0}\int\dfrac{d\omega}{2\pi}\left\langle \text{Tr}\left\{ v_{F}^{i}\mathbf{g}\right\} \right\rangle \label{eq:neutral-Eilenberger}
\end{align}
for the quasiparticle density and current density, 
\begin{align}
\rho_{e}\left(x\right) & =\mathbf{i}\pi e\hbar N_{0}\int\dfrac{d\omega}{2\pi}\left\langle \text{Tr}\left\{ \mathbf{g}\right\} \right\rangle \nonumber \\
j_{e}^{i}\left(x\right) & =\mathbf{i}\pi e\hbar N_{0}\int\dfrac{d\omega}{2\pi}\left\langle \text{Tr}\left\{ \tau_{3}v_{F}^{i}\mathbf{g}\right\} \right\rangle \label{eq:electric-Eilenberger}
\end{align}
for the electric charge and current densities, and 
\begin{align}
\rho_{s}^{k}\left(x\right) & =\mathbf{i}\pi\dfrac{g}{2}\hbar N_{0}\int\dfrac{d\omega}{2\pi}\left\langle \text{Tr}\left\{ \mathbf{s}_{k}\tau_{3}\mathbf{g}\right\} \right\rangle \nonumber \\
j_{s}^{ik}\left(x\right) & =\mathbf{i}\pi\dfrac{g}{2}\hbar N_{0}\int\dfrac{d\omega}{2\pi}\left\langle \text{Tr}\left\{ \mathbf{s}_{k}v_{F}^{i}\mathbf{g}\right\} \right\rangle \label{eq:spin-Eilenberger}
\end{align}
for the spin charge and current densities. We also have 
\begin{equation}
\Delta\left(x\right)=-\mathbf{i}\pi\hbar V_{0}N_{0}\int\dfrac{d\omega}{2\pi}\left\langle \text{Tr}\left\{ \mathbf{i}\sigma_{2}\tau_{+}\mathbf{g}\right\} \right\rangle \label{eq:self-consistency-Eilenberger}
\end{equation}
for the self-consistent relation of the gap parameter.

The Eilenberger equation \prettyref{eq:Eilenberger} constitutes a
convenient simplification in the description of the superconductor
phenomenology in the clean limit when $\tau\rightarrow\infty$. For
finite mean free time $\tau$, the self-consistency in $\left\langle \mathbf{g}\right\rangle $
might be problematic. Thanks to the normalization condition, one can
go further to the diffusive limit, called the Usadel limit.

\section{Usadel equation: diffusive limit\label{sec:Usadel-equation}}

The final approximation we will give in this paper is the diffusive
one, also called Usadel limit \cite{usadel.1970}. The associated
Usadel equation is a restriction of the Eilenberger one for diffusive
systems, when the self-consistent impurity contribution in \prettyref{eq:Eilenberger}
disapears in a resulting diffusion-like equation. For diffusive systems,
the $\xi$-integrated Green function can be expanded as 
\begin{equation}
\mathbf{g}=\mathbf{g}_{0}\left(\omega,x\right)+\boldsymbol{\hat{v}_{F}\cdot\hat{g}}\label{eq:Usadel-expansion}
\end{equation}
with an isotropic component $\mathbf{g}_{0}$ and a smaller contribution
$\boldsymbol{\hat{g}}$ along the Fermi velocity $\boldsymbol{\hat{v}_{F}}$
(the hat marks the unit vector). One then has $\left\langle \mathbf{g}\right\rangle =\mathbf{g}_{0}$.
The Usadel equation corresponds to the equation for the isotropic
part only. The derivation of the Usadel equation from the Eilenberger
one in the Abelian case is well described in \cite{usadel.1970,b.kopnin,Gorkov2008}
for instance, so we just sketch its generalisation below since there
is no more difficulty to deal with the non-Abelian situation.

The derivation of the Usadel equation relies on the normalization
condition \prettyref{eq:normalisation-condition}, which reads twofold
now: $\mathbf{g_{0}\cdot}\mathbf{g}_{0}=1$ and $\boldsymbol{\hat{g}}\cdot\mathbf{g}_{0}+\mathbf{g}_{0}\cdot\boldsymbol{\hat{g}}=\boldsymbol{0}$.
After multiplying the Eilenberger equation \prettyref{eq:Eilenberger}
with $\boldsymbol{\hat{v}_{F}}$ and averaging it as in \prettyref{eq:averaging-Fermi-surf},
one obtains
\begin{equation}
-\ell\mathbf{g}_{0}\boldsymbol{\mathfrak{D}}_{i}\mathbf{g}_{0}=\boldsymbol{\hat{g}}_{i}\label{eq:Usadel-substitution}
\end{equation}
after use of the normalization conditions several times, and with
$\ell=v_{F}\tau$ the mean free path. Next step is to average the
Eilenberger equation itself, and to substitute \prettyref{eq:Usadel-substitution}
into the resulting equation. One obtains then
\begin{multline}
\dfrac{\mathbf{i}\hbar c}{2}\left[\tau_{3},\boldsymbol{\mathfrak{D}}_{0}\mathbf{g}_{0}\right]_{+}-\mathbf{i}\hbar D\left(\boldsymbol{\mathfrak{D}}_{i}\mathbf{g}_{0}\right)\cdot\left(\boldsymbol{\mathfrak{D}}_{i}\mathbf{g}_{0}\right)\\
+\left[\hbar\omega\tau_{3}+\boldsymbol{\Delta},\mathbf{g}_{0}\right]_{-}=0\label{eq:Usadel}
\end{multline}
for the generalized Usadel equation in the presence of non-Abelian
gauge field, with $D=\ell v_{F}/3$ the diffusion constant.

Since the expressions \prettyref{eq:neutral-Eilenberger}, \prettyref{eq:electric-Eilenberger},
\prettyref{eq:spin-Eilenberger} and \prettyref{eq:self-consistency-Eilenberger}
already contains the averaging over the Fermi surface angular dependency,
it is sufficient to substitute the expansion \prettyref{eq:Usadel-expansion}
and the substitution \prettyref{eq:Usadel-substitution} to get
\begin{equation}
\Delta\left(x\right)=\mathbf{i}\pi\hbar V_{0}N_{0}\int\dfrac{d\omega}{2\pi}\text{Tr}\left\{ \mathbf{i}\sigma_{2}\tau_{+}\mathbf{g}_{0}\left(x,\omega\right)\right\} 
\end{equation}
for the self-consistency relation, 
\begin{align}
\rho_{n}\left(x\right) & =\mathbf{i}\pi\hbar N_{0}\int\dfrac{d\omega}{2\pi}\text{Tr}\left\{ \tau_{3}\mathbf{g}_{0}\left(\omega,x\right)\right\} \nonumber \\
j_{n}^{i}\left(x\right) & =-\mathbf{i}\pi D\hbar N_{0}\int\dfrac{d\omega}{2\pi}\text{Tr}\left\{ \mathbf{g}_{0}\boldsymbol{\mathfrak{D}}_{i}\mathbf{g}_{0}\right\} \label{eq:neutral-Usadel}
\end{align}
for the quasi-particle density and current, 
\begin{align}
\rho_{e}\left(x\right) & =\mathbf{i}\pi e\hbar N_{0}\int\dfrac{d\omega}{2\pi}\text{Tr}\left\{ \mathbf{g}_{0}\left(\omega,x\right)\right\} \nonumber \\
j_{e}^{i}\left(x\right) & =-\mathbf{i}\pi eD\hbar N_{0}\int\dfrac{d\omega}{2\pi}\text{Tr}\left\{ \tau_{3}\mathbf{g}_{0}\boldsymbol{\mathfrak{D}}_{i}\mathbf{g}_{0}\right\} \label{eq:electric-Usadel}
\end{align}
for the electric charge and current densities, and
\begin{align}
\rho_{e}\left(x\right) & =\mathbf{i}\pi\dfrac{g}{2}\hbar N_{0}\int\dfrac{d\omega}{2\pi}\text{Tr}\left\{ \mathbf{s}_{k}\tau_{3}\mathbf{g}_{0}\left(\omega,x\right)\right\} \nonumber \\
j_{e}^{ik}\left(x\right) & =-\mathbf{i}\pi\dfrac{g}{2}D\hbar N_{0}\int\dfrac{d\omega}{2\pi}\text{Tr}\left\{ \mathbf{s}_{k}\mathbf{g}_{0}\boldsymbol{\mathfrak{D}}_{i}\mathbf{g}_{0}\right\} \label{eq:spin-Usadel}
\end{align}
for the spin charge and densities, respectively. We see that the currents
contain magneto-electric contributions: the spin current contains
a term proportional to the electric charge, whereas the electric current
contains a term proportional to the spin charge, via the non-Abelian
covariant derivative \prettyref{eq:cov-deriv-super-2tensor} with
\prettyref{eq:A-def}.

\section{Poor-man derivation of the gauge-covariant Eilenberger equation\label{sec:Poor-man-Eilenberger}}

The two last sections of this paper contain extra materials, shortly
discussed. In this section, we discuss the derivation of the usual
Eilenberger equation using the so-called gradient expansion, and its
generalization to a gauge-covariant set of equations. We then recover
the non-Abelian Eilenberger equation \prettyref{eq:Eilenberger} in
a (perhaps) more direct way. In the next section, we use the result
of the present one to discuss the difference between a gauge potential
and a usual potential in term of transport equation. 

We thus suppose no gauge field for the moment. Then we define the
Wigner transformation as in \prettyref{eq:G-Wigner} and apply it
to the Dyson equation \prettyref{eq:Dyson-1} which then reads $\mathbf{G}^{-1}\left(p,x\right)\cdot e^{\hbar\Lambda/2\mathbf{i}}\cdot\mathbf{G}\left(p,x\right)=1$
with the Moyal operator $F\cdot\Lambda\cdot G=\partial_{p}F\cdot\partial_{x}G-\partial_{x}F\cdot\partial_{p}G$
for any functions $F$ and $G$, see \cite{Hillery1984,Polkovnikov2009}
for more details. At first order in a $\hbar$ expansion, one has
\begin{multline}
\mathbf{G}^{-1}\left(p,x\right)\mathbf{G}\left(p,x\right)\\
+\dfrac{\mathbf{i}\hbar}{2}\left(\partial_{p}\mathbf{G}^{-1}\cdot\partial_{x}\mathbf{G}-\partial_{x}\mathbf{G}^{-1}\cdot\partial_{p}\mathbf{G}\right)\approx1
\end{multline}
with $\mathbf{G}^{-1}\left(p,x\right)=\mathbf{G}_{0}^{-1}\left(p,x\right)+\boldsymbol{\Sigma}\left(p,x\right)$
in general, but we no more discuss the self-energy in the following.
We have $\mathbf{G}_{0}^{-1}\left(p,x\right)=\hbar\omega\tau_{3}-p^{2}/2m+\mu+\boldsymbol{\Delta}\left(x\right)$
for a conventional superconductor, with $\boldsymbol{\Delta}$ defined
in \prettyref{eq:Delta-bold}. Taking the difference of the Dyson
equation and its adjoint, one ends up with
\begin{equation}
\dfrac{\mathbf{i}\hbar}{2}\left[\tau_{3},\partial_{t}\mathbf{g}\right]_{+}+\mathbf{i}\hbar v_{F}^{i}\partial_{i}\mathbf{g}+\left[\hbar\omega\tau_{3}+\boldsymbol{\Delta},\mathbf{g}\right]_{-}=0\label{eq:Eilenberger-no-gauge}
\end{equation}
for the $\xi$-integrated Green functions in the vicinity of the Fermi
surface. We once again discard all the terms with momentum derivatives,
see \prettyref{sec:Eilenberger-equation}. The expression \prettyref{eq:Eilenberger-no-gauge}
is the so-called Eilenberger equation, when no gauge-field applies
\cite{eilenberger.1968,larkin_ovchinnikov.1969,b.kopnin}.

If one wants \prettyref{eq:Eilenberger-no-gauge} to become gauge-covariant
with respect to the gauge transformation $\mathbf{g}\rightsquigarrow\mathbf{R}\mathbf{g}\mathbf{R}^{\dagger}$,
one can promote the usual derivatives $\partial_{\mu}$ in \prettyref{eq:Eilenberger-no-gauge}
to some covariant derivatives \prettyref{eq:cov-deriv-super-2tensor}
which transforms as $\boldsymbol{\mathfrak{D}}_{\mu}\rightsquigarrow\mathbf{R}\boldsymbol{\mathfrak{D}}_{\mu}\mathbf{R}^{\dagger}$
when $\mathbf{A}_{\mu}\rightsquigarrow\mathbf{R}\mathbf{A}_{\mu}\mathbf{R}^{\dagger}-\mathbf{i}\mathbf{R}\partial_{\mu}\mathbf{R}^{\dagger}$.
Then we recover the gauge covariant Eilenberger equation \prettyref{eq:Eilenberger}
without the impurities corrections here for simplicity. Adding the
isotropic model for the scattering is straightforward, as well as
the derivation of the Usadel limit from there.

The above derivation is highly satisfying, since it does not require
the lengthy calculations of \prettyref{sec:Transport-equations} and
\prettyref{sec:Eilenberger-equation} to obtain the gauge-covariant
Eilenberger equation. Nevertheless, the general transport equation
\prettyref{eq:transport-diff} cannot be obtained using a simple argument
of covariance, since the gauge fields are present there.

\section{The exchange field as a usual potential\label{sec:Potential-or-gauge-potential}}

One can really wonder whether it makes sense or not to discuss a complicated
gauge theory to include magnetic interaction. Indeed, conventional
ferromagnetism seems to be properly described when adding the Zeeman
term $h_{Z}\sigma_{3}$ into the equation of motion as a regular potential,
not a gauge potential. Here I clarify a bit the difference between
the two approaches.

When writing the propagator
\begin{multline}
\mathbf{G}_{0}^{-1}\left(x\right)=\mathbf{i}\hbar c\tau_{3}\mathbf{D}_{0}+\dfrac{\hbar^{2}}{2m}\boldsymbol{\partial_{x}\cdot\partial_{x}}+\boldsymbol{\Delta}\left(x,t\right)\\
=\mathbf{i}\hbar\tau_{3}\partial_{t}-\hbar\tau_{3}\mathbf{A}_{0}\left(x,t\right)+\dfrac{\hbar^{2}}{2m}\boldsymbol{\partial_{x}\cdot\partial_{x}}+\boldsymbol{\Delta}\left(x,t\right)\label{eq:G-1-potential}
\end{multline}
one has the choice to express 
\begin{equation}
\mathbf{A}_{0}=\tau_{3}\dfrac{h_{Z}\sigma_{3}-\mu}{\hbar}\label{eq:Zeeman-gauge-potential}
\end{equation}
either in term of a gauge-potential (first line of \prettyref{eq:G-1-potential})
or as a usual potential (second line of \prettyref{eq:G-1-potential}).
Reproducing the derivation in the previous section, we find (see \cite{golubov_kupriyanov.2004,buzdin.2005_RMP,bergeret_volkov_efetov_R.2005}
for a discussion of the consequences of this equation)
\begin{equation}
\dfrac{\mathbf{i}\hbar}{2}\left[\tau_{3},\partial_{t}\bar{\mathbf{g}}\right]_{+}+\mathbf{i}\hbar v_{F}^{i}\partial_{i}\bar{\mathbf{g}}+\left[\hbar\omega\tau_{3}-h_{Z}\sigma_{3}+\boldsymbol{\Delta},\bar{\mathbf{g}}\right]_{-}=0\label{eq:Eilenberger-Zeeman}
\end{equation}
when we suppose \prettyref{eq:Zeeman-gauge-potential} to be a usual
potential. In contrary, the gauge covariant Eilenberger equation \prettyref{eq:Eilenberger}
leads to 
\begin{multline}
\dfrac{\mathbf{i}\hbar}{2}\left[\tau_{3},\partial_{t}\tilde{\mathbf{g}}+\dfrac{\mathbf{i}}{\hbar}\left[h_{Z}\tau_{3}\sigma_{3},\tilde{\mathbf{g}}\right]_{-}\right]_{+}+\mathbf{i}\hbar v_{F}^{i}\partial_{i}\tilde{\mathbf{g}}\\
+\left[\hbar\omega\tau_{3}+\boldsymbol{\Delta},\tilde{\mathbf{g}}\right]_{-}=0\label{eq:Eilenberger-Zeeman-covariant}
\end{multline}
where the time-covariant-derivative is displayed explicitly. 

The two equations \prettyref{eq:Eilenberger-Zeeman} and \prettyref{eq:Eilenberger-Zeeman-covariant}
do not look the same, and questions rise up about the correctness
of the present approach. To resolve this paradox, note that the two
functions $\mathbf{g}$ in \prettyref{eq:Eilenberger-Zeeman} and
\prettyref{eq:Eilenberger-Zeeman-covariant} are not equivalent: $\bar{\mathbf{g}}$
in \prettyref{eq:Eilenberger-Zeeman} corresponds to the Wigner transformed
correlation function adapted to a situation without gauge field \prettyref{eq:G-Wigner}
(more explicitly, $\bar{\mathbf{g}}$ is the $\xi$-integrated, mixed-Fourier
transform of the Green-Gor'kov functions $\mathbf{G}\left(x_{1},x_{2}\right)$),
whereas $\tilde{\mathbf{g}}$ in \prettyref{eq:Eilenberger-Zeeman-covariant}
is the result of the gauge-covariant Wigner transformation with curvature
\prettyref{eq:G-Wigner-UGU}. In addition, the gauge field associated
to the gauge potential \prettyref{eq:Zeeman-gauge-potential} is rather
trivial $\mathbf{F}_{i0}=\tau_{3}\left(\sigma_{3}\partial_{i}h_{Z}-\partial_{i}\mu\right)/\hbar$
and contains only the non-Abelian generalization of the electric field.
This is obvious since the electric field is the only one having a
pure potential contribution. This suggests that one may possibly kill
the gauge potential in \prettyref{eq:Eilenberger-Zeeman-covariant}.
Indeed, it is always possible to work in a gauge such that the time-sector
of the gauge potential $\mathbf{A}_{0}=0$ vanishes, called the temporal,
or Weyl gauge \cite{Coleman1988}. 

For simplicity we assume in the following that the exchange field
$h_{Z}$ is space-time independent, in which case the non-Abelian
gauge field disappears. Then the gauge potential can be canceled explicitly
in \prettyref{eq:Eilenberger-Zeeman-covariant} using the gauge transformation
$\mathbf{R}_{h}=\exp\left[-\mathbf{i}\tau_{3}\sigma_{3}h_{Z}t/\hbar\right]$.
Transforming $\tilde{\mathbf{g}}=\mathbf{R}_{h}\tilde{\mathbf{g}}'\mathbf{R}_{h}^{\dagger}$
reduces \prettyref{eq:Eilenberger-Zeeman-covariant} to \prettyref{eq:Eilenberger-no-gauge}
for the $\tilde{\mathbf{g}}'$ functions, when no gauge field is present.
In particular, the transformation does not alter the gap parameter,
since it has a singlet representation in the spin sector. We note
that the transformation looks like the replacement $\omega'=\omega-h_{Z}\sigma_{3}\tau_{3}/\hbar$
for $\tilde{\mathbf{g}}$, to be compared with the expression \prettyref{eq:Eilenberger-Zeeman}. 

This intriguing result is partially consistent with the old-known
result that the gap parameter is unaffected by a weak paramagnetic
interaction established by Sarma \cite{Sarma1963}. Sarma invoked
the singlet symmetry of the gap parameter as well to understand his
result; we simply recast this argument into a gauge invariance in
\prettyref{sec:Green-Gorkov-equation}. Nevertheless, Sarma also found
that a large exchange field $h_{Z}>\Delta$ alters the critical line
transition. In the Eilenberger formalism, the high energy sector $h_{Z}>\Delta$
is in principle not reachable, but the gauge transformation $\mathbf{R}_{h}$
discards the constant paramagnetic interaction for all energies, since
it can be applied to the Dyson-Gor'kov's equation (\prettyref{sec:Green-Gorkov-equation})
as well. It is not yet clear whether the Sarma's result can be entirely
understood or not in the gauge formalism I propose here. It might
well be that the mean field treatment of the superconducting phase
imposes some restriction on the use of the gauge redundancy. It is
for instance clear that a large enough exchange field compensates
the kinetic energy of the quasi-particles. Perhaps the gauge invariance
of the interaction Hamiltonian \prettyref{eq:H-int} is verified only
at low energies, at least for energies smaller than the gap parameter.
A detailed study of this effect is postponed to future works, but
I fear a complete discussion of the pair-destruction mechanism requires
the self-consistent treatment of the Cooper instability mechanism:
an effective theory with coupling constants as discussed in this paper
might well be not powerful enough. 

Thus, expressions \prettyref{eq:Eilenberger-Zeeman} and \prettyref{eq:Eilenberger-Zeeman-covariant}
lead to the same conclusion in the low energy sector $\Delta<h_{Z}$
and for constant exchange field at zero temperature, according to
the Sarma's result \cite{Sarma1963}. For a space-dependent exchange
field, the cancelation of the exchange field is not a trivial task,
and further discussions are necessary. Also, generalizations to non-zero-temperatures
should be done with care, because the Matsubara formalism alters the
gauge structure of the theory, see \textit{e.g.} \cite{Zinn-Justin2002}. 

Obviously a spin-orbit term can not be canceled by a gauge transformation
affecting the time-sector only. Additionally, the choice of the temporal
gauge will not alter the spin-orbit interaction. Then, to understand
the interaction between (at least) Zeeman and spin-orbit interaction,
the gauge formalism I developed in this study should be useful.

\textit{Note added in proof:} An other way to treat the spin interaction
is to consider the spin-orbit interaction as a gauge potential, whereas
the exchange field is treated as a conventional potential. This method,
especially useful in the case of stationary problems, leads then to
the Eilenberger equation
\begin{equation}
\mathbf{i}\hbar v_{F}^{i}\mathfrak{D}_{i}\mathbf{g}+\left[\hbar\omega\tau_{3}-h_{Z}\sigma_{3}+\boldsymbol{\Delta},\mathbf{g}\right]_{-}=0
\end{equation}
and can be conveniently transformed toward the Matsubara formalism
using a Wick rotation \cite{b.abrikosov_gorkov}. Additionally, using
a gauge-covariant Wigner transformation for the space coordinates
only allows to include non-stationary effect via the Keldysh space.
This last approach has been recently followed by Bergeret and Tokatly
\cite{Bergeret2014}, who derived essentially the same equations as
in this paper, using the same covariant method but transforming only
the space components of the Green functions. In the notations of the
present paper, they thus transform the Green functions from $\mathbf{G}\left(\boldsymbol{x}_{1},\boldsymbol{x}_{2},t_{1},t_{2}\right)$
to $\mathbf{G}\left(\boldsymbol{p},\boldsymbol{x},t_{1},t_{2}\right)$
using $\mathbf{G}\left(\boldsymbol{p},\boldsymbol{x},t_{1},t_{2}\right)=\int d\boldsymbol{z}\left[e^{-\mathbf{i}\boldsymbol{p\cdot z}/\hbar}\mathbf{U}\left(\boldsymbol{x},\boldsymbol{x}_{1}\right)\mathbf{G}\left(\boldsymbol{x}_{1},\boldsymbol{x}_{2},t_{1},t_{2}\right)\mathbf{U}\left(\boldsymbol{x}_{2},\boldsymbol{x}\right)\right]$
as the gauge-covariant Wigner transformation, the main difference
being that $\mathbf{G}\left(\boldsymbol{x}_{1},\boldsymbol{x}_{2},t_{1},t_{2}\right)$
then stands for some matrix in the Keldysh plus particle-hole plus
spin plus charge space, a complication I wanted to avoid here. They
do not discuss non-stationary problems, though. At the time of writing,
it is not clear which of these approaches (to treat the exchange field
as the time-sector of the gauge-potential or not, and/or to Wigner-transform
the time variables of the Green functions or not) will get the more
tractable analysis of relevant situations. For a comprehensible review
of the difficulties to deal with non-equilibirum superconductors in
the quasi-classical limit, one can consult \cite{b.kopnin,Eschrig2009a}
and references therein.

\section{Conclusion and Perspectives\label{sec:Conclusion-and-Perspectives}}

In this study, I focused on the establishment of a family of transport-like
equations which are of possible interest for the study of superconductors
in the presence of magnetic interaction like space-time dependent
Zeeman and/or spin-orbit interaction linear in the momentum. Having
in mind the recently discussed spin texture competing with the superconducting
order, I proposed to enlarge the usual description of the electromagnetic
interaction in a gauge interaction including non-Abelian spin plus
Abelian charge sectors. I show how the Gor'kov set of equations can
be generalized to a $U\left(1\right)\otimes SU\left(2\right)\otimes SU\left(2\right)$,
charge plus spin plus particle-hole gauge theory (\prettyref{sec:Green-Gorkov-equation}). 

Thanks to the well established gauge principles, the proposed description
takes into account the self-consistent interactions between the gauge
degrees of freedom and the superconducting phase (\prettyref{sec:Definition-and-Convention}). 

This set of equations is nevertheless intrinsically non-linear and
self-consistent. To simplify it, I thus proposed to reduce the quantum
structure of the Dyson-Gor'kov equations toward a transport-like theory
at the quasi-classical level, when the quasi-classical Green function
$\mathbf{G}\left(p,x\right)$ now describes the normal and anomalous
correlation functions in a time-dependent phase-space (\prettyref{sec:Transport-equations},
in particular expressions \prettyref{eq:transport-diff} and \prettyref{eq:transport-sum}). 

In addition, the superconducting state usually has a clear energy
scale separation $\Delta/E_{F}\ll1$ between the gap parameter $\Delta$
and the Fermi energy $E_{F}$, which allows to reduce even further
the transport equation into the so-called Eilenberger equation, here
generalized to take into account the electronic spin degree of freedom
on the same footing as the charge one (\prettyref{sec:Eilenberger-equation},
especially \prettyref{eq:Eilenberger}). The diffusive limit of the
Eilenberger equation, known as the Usadel equation, is also given
(\prettyref{sec:Usadel-equation}, see \prettyref{eq:Usadel}). In
each case (quasi-classical transport, Eilenberger and Usadel) I provided
the associated quasi-particle, electric and spin charges currents.
In particular, the charge and the spin currents now exhibit some magneto-electric
couplings, which will be discussed in subsequent studies. 

The different levels of approximations discussed in this paper may
constitute an interesting way for studying the topological problems
in condensed matter, when Zeeman plus spin-orbit effects compete with
the superconducting proximity effect in disordered structures, as
well as to address fundamental questions in bulk magnetic superconductors.
Since the equations I derived contain the limiting cases of normal
metal interacting or not with a (non-)Abelian gauge field, this study
may be of interest for the understanding of quantum Hall effects,
for spintronics, and for many more topics in condensed matter when
the internal degrees of freedom of the electrons need to be correctly
accounted for. In any case, this study should pave the way toward
a better understanding of the relations between condensed matter and
gauge theories.

Among other urgent problems to be resolved in the gauge formalism
proposed in this study are the inclusion of boundary conditions in
the transport formalism, the discussion of the lattice symmetry in
addition to the gauge redundancy, and the understanding of the gauge
properties of the superconducting phase in a statistical field theory.
It should be interesting to understand the role of a possible quantization
of the gauge field on the superconducting phase as well, and its associated
phenomenology (anomaly, confinement, ...). The intrinsic non-linearity
of the non-Abelian formalism suggests that the proposed transport-like
theory exhibit some sort of instantons, too. This has to be checked
as well. In mesoscopic physics terms, the proposed formalism opens
the way to discuss the role of impurities in a self-consistent way.
Possible applications are in the recently emerging field of topological
matter and its relation to quantum information perspectives.
\begin{acknowledgments}
Stimulating discussions with C. Gorini were particularly appreciated
during the Annual Meeting of the French Mesoscopic Physics GDR in
Aussois, December 2013. Discussions with F. S. Bergeret and I. V.
Tokatly about their work were particularly enjoyed, too. I thank F.
Hassler and G. Viola for daily helpful remarks, as well as B. Bergeret,
G. Catelani, B. Douçot, C. Ohm and J. Ulrich for their interests in
this work. I am grateful for support from the Alexander von Humboldt
foundation.
\end{acknowledgments}
\appendix

\section{Correspondence rules\label{sec:ANNEX-Correspondence-rules}}

In this appendix, we discuss the generic rules of transformation from
the Dyson equation to some quantum transport equations through the
gauge-covariant Wigner transformation \prettyref{eq:G-Wigner-UGU}.
In particular, we sum up the long algebra required to obtain the first
and second order covariant derivatives of the gauge-covariant Wigner-transformation
introduced in the main text. The equations below are generic, and
can be applied to any gauge-field $F_{\mu\nu}$ defined from a gauge
potential $A_{\mu}$ through the definition $F_{\mu\nu}=\partial_{\mu}A_{\nu}-\partial_{\nu}A_{\mu}+\mathbf{i}\left[A_{\mu},A_{\nu}\right]$.
We intensively use the relativistic tensorial notations in this appendix;
they are sum-up at the beginning of \prettyref{sec:Green-Gorkov-equation}.
The covariant derivatives are defined as $D_{\mu}=\partial_{\mu}+\mathbf{i}A_{\mu}$
and $D_{\mu}^{\dagger}=\partial_{\mu}-\mathbf{i}A_{\mu}$ where the
derivative applies to the left in $D_{\mu}^{\dagger}$ and $A_{\mu}$
is a Hermitian field. Finally the covariant derivative is $\mathfrak{D}^{\mu}\left(x\right)F_{\mu\nu}\left(x\right)=\partial F_{\mu\nu}/\partial x_{\mu}+\mathbf{i}\left[A^{\mu}\left(x\right),F_{\mu\nu}\left(x\right)\right]$
when applied to the second-rank gauge-field tensor. In the following,
both the quasi-classical Green function $G\left(p,x\right)$ and the
gauge field $F_{\mu\nu}\left(x\right)$ behave as second-rank tensors. 

One then defines the geometric differential propagator, applied to
a field amplitude or a second-rank tensor $F\left(x\right)$ (demonstrated
in \cite{Elze1986} using an expansion of the exponential)

\begin{equation}
e^{y\cdot\mathfrak{D}\left(x\right)}F\left(x\right)=U\left(x,x+y\right)F\left(x+y\right)U\left(x+y,x\right)
\end{equation}
with the path-ordered integral $U\left(x,y\right)$ defined in \prettyref{eq:U-DEF},
also called parallel displacement operator, Wilson line or link operator
\cite{Itzykson2006}. The parallel transport operator is an operator
in gauge-space only (for instance, in the charge $\otimes$ spin-space,
or the Nambu $\otimes$ charge $\otimes$ spin-space in the main text),
whereas it is just a phase-shift in real-space, as can be seen from
its property \prettyref{eq:U-Psi}. The last expressions we need to
proceed are the covariant derivatives of the parallel shift operator

\begin{multline}
D_{\nu}\left(b\right)U\left(b,a\right)=\mathbf{i}\left(b-a\right)^{\mu}\times\\
\int_{0}^{1}ds\left[sU\left(b,\tau_{s}\right)F_{\mu\nu}\left(\tau_{s}\right)U\left(\tau_{s},a\right)\right]
\end{multline}
\begin{multline}
U\left(b,a\right)D_{\nu}^{\dagger}\left(a\right)=\mathbf{i}\left(b-a\right)^{\mu}\\
\int_{0}^{1}ds\left[\left(1-s\right)U\left(b,\tau_{s}\right)F_{\mu\nu}\left(\tau_{s}\right)U\left(\tau_{s},a\right)\right]\label{eq:DU}
\end{multline}
the demonstration of which are in \cite{Elze1986,Winter1984}, and
where $\tau_{s}=a+\left(b-a\right)s$ represents the straight line
between the extremum points $a$ and $b$, as discussed in \cite{Elze1986,Vasak1987}.
They include non-trivial curvature effect due to the presence of the
gauge field in their right-hand-side. We define the gauge-covariant
Wigner-transform \prettyref{eq:G-Wigner-DEF} adapting the treatment
given by Elze, Gyulassy and Vasak \cite{Elze1986} to the Green function
problem. We then use the property \prettyref{eq:U-Psi} to write the
workable representation of the Wigner transformation in \prettyref{eq:G-Wigner-UGU}.

The change of coordinates from the two-points correlators and the
phase-space quasi-classical Green function reads $x_{1,2}=x\mp z/2$
and $\partial_{1,2}=\partial_{x}/2\mp\partial_{z}$. Then one obtains
-- be warn that the notations are mixed in the formulas below (they
should display only the variables $x$ and $z$) for the sake of notational
compactness --
\begin{multline}
\dfrac{\partial}{\partial x_{1}^{\nu}}U\left(\dfrac{x_{1}+x_{2}}{2},x_{1}\right)=\mathbf{i}U\left(x,x_{1}\right)A_{\nu}\left(x_{1}\right)\\
-\dfrac{\mathbf{i}}{2}A_{\nu}\left(x\right)U\left(x,x_{1}\right)+\dfrac{\mathbf{i}}{2}\int_{0}^{1}ds\left[\mathfrak{F}_{\nu}^{s-1}\left(x,z\right)\right]U\left(x,x_{1}\right)\label{eq:appendix-d1U1}
\end{multline}
\begin{multline}
\dfrac{\partial}{\partial x_{1}^{\nu}}U\left(x_{2},\dfrac{x_{1}+x_{2}}{2}\right)=\dfrac{\mathbf{i}}{2}U\left(x_{2},x\right)A_{\nu}\left(x\right)\\
+\dfrac{\mathbf{i}}{2}U\left(x_{2},x\right)\int_{0}^{1}ds\left[\mathfrak{F}_{\nu}^{s}\left(x,z\right)\right]\label{eq:appendix-d1U2}
\end{multline}
\begin{multline}
\dfrac{\partial}{\partial x_{2}^{\nu}}U\left(\dfrac{x_{1}+x_{2}}{2},x_{1}\right)=-\dfrac{\mathbf{i}}{2}A_{\nu}\left(x\right)U\left(x,x_{1}\right)\\
+\dfrac{\mathbf{i}}{2}\int_{0}^{1}ds\left[\bar{\mathfrak{F}}_{\nu}^{s-1}\left(x,z\right)\right]U\left(x,x_{1}\right)
\end{multline}
\begin{multline}
\dfrac{\partial}{\partial x_{2}^{\nu}}U\left(x_{2},\dfrac{x_{1}+x_{2}}{2}\right)=-\mathbf{i}A_{\nu}\left(x_{2}\right)U\left(x_{2},x\right)\\
+\dfrac{\mathbf{i}}{2}U\left(x_{2},x\right)A_{\nu}\left(x\right)+\dfrac{\mathbf{i}}{2}U\left(x_{2},x\right)\int_{0}^{1}ds\left[\bar{\mathfrak{F}}_{\nu}^{s}\left(x,z\right)\right]
\end{multline}
where we used that the path is a straight line, so we can write some
expressions like $U\left(z,x_{1}\right)=U\left(z,x_{1}\right)U\left(x_{1},x\right)U\left(x,x_{1}\right)=U\left(z,x\right)U\left(x,x_{1}\right)$
since $U\left(x_{1},x\right)U\left(x,x_{1}\right)=1$ is not a Wilson
loop, in which case it might be a phase factor. The same applies for
$U\left(x_{2},z\right)=U\left(x_{2},x\right)U\left(x,z\right)$, which
allows the expressions to be written in terms of the $\mathfrak{D}\left(x\right)$
operator (see also \prettyref{eq:F-integral} below). We propose the
notational simplifications
\[
\mathfrak{F}_{\nu}^{s}\left(x,z\right)=\dfrac{1-s}{2}e^{sz\cdot\mathfrak{D}\left(x\right)/2}z^{\mu}F_{\mu\nu}\left(x\right)
\]
\begin{eqnarray}
\bar{\mathfrak{F}}_{\nu}^{s}\left(x,z\right) & = & \dfrac{1+s}{2}e^{sz\cdot\mathfrak{D}\left(x\right)/2}z^{\mu}F_{\mu\nu}\left(x\right)\label{eq:F-integral}
\end{eqnarray}
which keep the following calculations tractable. Essentially, passing
from $\mathfrak{F}$ to $\bar{\mathfrak{F}}$ consists in changing,
in the associated integral, the direction of propagation from the
center-of-mass coordinate $x$ to one of the extremum $x_{1}$ or
$x_{2}$ along a straight line. 

We can now evaluate
\begin{multline*}
U\left(x,x_{1}\right)\left[D_{\nu}\left(x_{1}\right)G\left(x_{1},x_{2}\right)\right]U\left(x_{2},x\right)=\\
\dfrac{\partial}{\partial x_{1}^{\nu}}\tilde{G}+\dfrac{\mathbf{i}}{2}\left[A_{\nu}\left(x\right),\tilde{G}\right]\\
-\dfrac{\mathbf{i}}{2}\int_{0}^{1}ds\left[\mathfrak{F}_{\nu}^{s-1}\left(x,z\right)\tilde{G}+\tilde{G}\mathfrak{F}_{\nu}^{s}\left(x,z\right)\right]
\end{multline*}

\begin{multline}
U\left(x,x_{1}\right)\left[G\left(x_{1},x_{2}\right)D_{\nu}^{\dagger}\left(x_{2}\right)\right]U\left(x_{2},x\right)=\\
\dfrac{\partial}{\partial x_{2}^{\nu}}\tilde{G}+\dfrac{\mathbf{i}}{2}\left[A_{\nu}\left(x\right),\tilde{G}\right]\\
-\dfrac{\mathbf{i}}{2}\int_{0}^{1}ds\left[\mathscr{\bar{\mathfrak{F}}}_{\nu}^{s-1}\left(x,z\right)\tilde{G}+\tilde{G}\bar{\mathscr{\mathfrak{F}}}_{\nu}^{s}\left(x,z\right)\right]\label{eq:covariant-derivative-Wigner-integrand}
\end{multline}
with $\tilde{G}=U\left(x,x_{1}\right)G\left(x_{1},x_{2}\right)U\left(x_{2},x\right)$.
Finally, one can calculate easily the gauge-covariant Wigner-transformation
of the covariant derivative of the Green function as the Fourier transform
of the previous expressions. It gives:

\begin{multline}
\int dz\left[e^{-\mathbf{i}p\cdot z/\hbar}U\left[D_{\nu}\left(x_{1}\right)G\left(x_{1},x_{2}\right)\right]U\right]=\\
\left(\dfrac{1}{2}\mathfrak{D}_{\nu}\left(x\right)-\mathbf{i}\dfrac{p_{\nu}}{\hbar}\right)G\left(p,x\right)\\
-\dfrac{\mathbf{i}}{2}\left\{ \left[\int_{0}^{1}ds\mathfrak{F}_{\nu}^{s-1}\right]G\left(p,x\right)+G\left(p,x\right)\left[\int_{0}^{1}ds\mathfrak{F}_{\nu}^{s}\right]\right\} \label{eq:D-Wigner-1}
\end{multline}

\begin{multline}
\int dz\left[e^{-\mathbf{i}p\cdot z/\hbar}U\left[G\left(x_{1},x_{2}\right)D_{\nu}^{\dagger}\left(x_{2}\right)\right]U\right]=\\
\left(\dfrac{1}{2}\mathfrak{D}_{\nu}\left(x\right)+\mathbf{i}\dfrac{p_{\nu}}{\hbar}\right)G\left(p,x\right)\\
-\dfrac{\mathbf{i}}{2}\left\{ \left[\int_{0}^{1}ds\bar{\mathscr{\mathfrak{F}}}_{\nu}^{s-1}\right]G\left(p,x\right)+G\left(p,x\right)\left[\int_{0}^{1}ds\bar{\mathfrak{F}}_{\nu}^{s}\right]\right\} \label{eq:D-Wigner-2}
\end{multline}
using some integration by part of the $\partial_{z}$ term, and the
symbolic formula $\int dz\left[e^{-\mathbf{i}p\cdot z/\hbar}f\left(z\right)g\left(z\right)\right]=f\left(\mathbf{i}\hbar\partial_{p}\right)\int dz\left[e^{-\mathbf{i}p\cdot z/\hbar}g\left(z\right)\right]$,
so that all the contributions $\mathfrak{F}$ have to be understood
as being $\mathfrak{F}\left(x,\mathbf{i}\hbar\partial_{p}\right)$
dependent. At the end, only the Wigner-Green-function $G\left(p,x\right)$
depends on $p$, so the momentum derivatives apply to $G\left(p,x\right)$
only. 

To calculate the gauge-covariant Wigner-transform of the second order
covariant derivative of the Green function, a convenient method is
to rewrite
\begin{multline}
U\left(x,x_{1}\right)\left[D_{\nu}\left(x_{1}\right)D^{\nu}\left(x_{1}\right)G\left(x_{1},x_{2}\right)\right]U\left(x_{2},x\right)=\\
U\left(\partial_{\nu}\left(x_{1}\right)D^{\nu}G\right)U+\mathbf{i}UA_{\nu}\left(x_{1}\right)\left(D^{\nu}G\right)U\\
=\partial_{\nu}\left[U\left(D^{\nu}G\right)U\right]-\left(\partial_{\nu}U\right)\left(D^{\nu}G\right)U\\
-U\left(D^{\nu}G\right)\left(\partial_{\nu}U\right)+\mathbf{i}UA_{\nu}\left(x_{1}\right)\left(D^{\nu}G\right)U\label{eq:expansion-second-order}
\end{multline}
where we do not write explicitly all the coordinates on the right-hand-side
when they are trivially reproduced from the left-hand-side. All the
derivatives are with respect to the first argument $x_{1}$ of the
Green function $G\left(x_{1},x_{2}\right)$. Then, we use the formula
\prettyref{eq:appendix-d1U1} and \prettyref{eq:appendix-d1U2} such
that the last term of \prettyref{eq:expansion-second-order} disappears
with the first term of \prettyref{eq:appendix-d1U1} and we are left
with
\begin{multline}
U\left(x,x_{1}\right)\left[D_{\nu}\left(x_{1}\right)D^{\nu}\left(x_{1}\right)G\left(x_{1},x_{2}\right)\right]U\left(x_{2},x\right)=\\
\partial_{\nu}\left[U\left(D^{\nu}G\right)U\right]+\dfrac{\mathbf{i}}{2}\left[U\left(D^{\nu}G\right)U,A_{\nu}\left(x\right)\right]\\
-\dfrac{\mathbf{i}}{2}\left\{ \int ds\left[\mathfrak{F}_{\nu}^{s-1}\right]U\left(D^{\nu}G\right)U+U\left(D^{\nu}G\right)U\int ds\left[\mathfrak{F}_{\nu}^{s}\right]\right\} 
\end{multline}
then, we just have to evaluate the derivative of \prettyref{eq:covariant-derivative-Wigner-integrand}
and to produce some algebra. Note the group structure of the covariant
derivative, since the above equation is exactly the same as \prettyref{eq:covariant-derivative-Wigner-integrand}
when we replace $UGU=\tilde{G}$ by $U\left(DG\right)U$. After tedious
algebra, one obtains
\begin{multline}
\int dz\left[e^{-\mathbf{i}p\cdot z/\hbar}U\left[D_{\nu}D^{\nu}\left(x_{1}\right)G\left(x_{1},x_{2}\right)\right]U\right]=\\
\left(\dfrac{1}{2}\mathfrak{D}_{\nu}\left(x\right)-\mathbf{i}\dfrac{p_{\nu}}{\hbar}\right)\left(\dfrac{1}{2}\mathfrak{D}^{\nu}\left(x\right)-\mathbf{i}\dfrac{p^{\nu}}{\hbar}\right)G\left(p,x\right)\\
-\mathbf{i}\int_{0}^{1}ds\left[\mathfrak{F}_{\nu}^{s-1}\right]\left(\dfrac{1}{2}\mathfrak{D}^{\nu}\left(x\right)-\mathbf{i}\dfrac{p^{\nu}}{\hbar}\right)G\left(p,x\right)\\
-\mathbf{i}\left(\dfrac{1}{2}\mathfrak{D}^{\nu}\left(x\right)-\mathbf{i}\dfrac{p^{\nu}}{\hbar}\right)G\left(p,x\right)\int_{0}^{1}ds\left[\mathfrak{F}_{\nu}^{s}\right]\\
-\dfrac{\mathbf{i}}{2}\int_{0}^{1}ds\left[\dfrac{2-s}{2}\mathfrak{D}^{\nu}\left(x\right)\mathfrak{F}_{\nu}^{s-1}\right]G\left(p,x\right)\\
-\dfrac{\mathbf{i}}{2}G\left(p,x\right)\int_{0}^{1}ds\left[\dfrac{1-s}{2}\mathfrak{D}^{\nu}\left(x\right)\mathfrak{F}_{\nu}^{s}\right]\\
+\dfrac{1}{4}\int_{0}^{1}ds\int_{0}^{1}d\tilde{s}\left(1-s\right)\left[\mathfrak{F}_{\left(1-s\right)\left(1-\tilde{s}\right)}^{\nu}\mathfrak{F}_{\nu}^{s-1}\right.\\
\left.-\mathfrak{F}_{\nu}^{s-1}\mathfrak{F}_{\tilde{s}\left(s-1\right)}^{\nu}\right]G\left(p,x\right)\\
+\dfrac{1}{4}G\left(p,x\right)\int_{0}^{1}ds\int_{0}^{1}d\tilde{s}s\left[\mathfrak{F}_{\nu}^{s}\mathfrak{F}_{\tilde{s}s}^{\nu}-\mathfrak{F}_{s\left(1-\tilde{s}\right)}^{\nu}\mathfrak{F}_{\nu}^{s}\right]\\
-\dfrac{1}{4}\left\{ \left[\int_{0}^{1}ds\mathfrak{F}_{s-1}^{\nu}\right]^{2}G\left(p,x\right)+G\left(p,x\right)\left[\int_{0}^{1}ds\mathfrak{F}_{s}^{\nu}\right]^{2}\right\} \\
-\dfrac{1}{2}\left\{ \int_{0}^{1}ds\left[\mathfrak{F}_{s-1}^{\nu}\right]G\left(p,x\right)\int_{0}^{1}ds\left[\mathfrak{F}_{\nu}^{s}\right]\right\} \label{eq:DD_Wigner_transform}
\end{multline}
for the gauge-covariant Wigner-transformation of the second covariant
derivative of the Green function, where all the $\mathfrak{F}$ contributions
are formal formula, to be understood as $\mathfrak{F}\left(x,\mathbf{i}\hbar\partial_{p}\right)$.
We used the symbolic notation
\begin{equation}
\mathfrak{D}^{\nu}\left(x\right)\mathfrak{F}_{\nu}^{s}\left(x,z\right)=\dfrac{1-s}{2}e^{sz\cdot\mathfrak{D}\left(x\right)/2}z^{\mu}\mathfrak{D}^{\nu}\left(x\right)F_{\mu\nu}\left(x\right)
\end{equation}
for notational convenience. Note that $s$ and $\tilde{s}$ are the
path arguments of the $\mathfrak{F}$ functional and their co- or
contra-variant position is meaningless, \textit{i.e.} $\mathfrak{F}^{s}=\mathfrak{F}_{s}$.
In contrary $\nu$ is the component of the gauge field strength tensor,
so $\mathfrak{F}_{\nu}\neq\mathfrak{F}^{\nu}=g^{\nu\mu}\mathfrak{F}_{\mu}$.
Note that the notations could be confusing: to adapt \prettyref{eq:DD_Wigner_transform}
to the main text, the greek indices $\nu$ should be replaced by some
latin ones. This is because only the space Laplacian appears in the
Dyson equations \prettyref{eq:Dyson-1} and \prettyref{eq:Dyson-2}.

The same calculation as before for the transformation of the second
order covariant derivative applied to the second variable $x_{2}$
gives the same expression \prettyref{eq:DD_Wigner_transform} except
for the replacement $p_{\nu}\rightarrow-p_{\nu}$ and $\mathfrak{F}\rightarrow\bar{\mathfrak{F}}$
since the derivative applies to $x_{2}=x+z/2$ (these are the same
differences as between \prettyref{eq:D-Wigner-1} and \prettyref{eq:D-Wigner-2}
obtained for the first order covariant derivatives).

The method explained above to get the Wigner transformation of the
covariant derivative can in principle be continued to the third order
covariant derivative, and so on. Nevertheless, having in mind the
free particle model of the main text, we here stop at the second order
derivative.

To conclude this set of rules for the transformation toward the transport
equation of the Green-Wigner function $G\left(p,x\right)$, we give
the transformation rule of a scalar potential, independent of the
impulsion, which reads $V\left(x_{1}\right)G\left(x_{1},x_{2}\right)$
in the Dyson equation. Then we apply the rule
\begin{multline}
\int dz\left[e^{-\mathbf{i}p\cdot z/\hbar}UV\left(x_{1}\right)G\left(x_{1},x_{2}\right)U\right]\\
=V\left(x-\mathbf{i}\hbar\partial_{p}/2\right)G\left(p,x\right)\label{eq:VG-exact}
\end{multline}
since the potential is a scalar, it commutes with the parallel displacement
operator, which acts only in the Lie algebra sub-space. We supposed
that the potential can be expanded in series, as usual. The same argument
gives
\begin{multline}
\int dz\left[e^{-\mathbf{i}p\cdot z/\hbar}UG\left(x_{1},x_{2}\right)V\left(x_{2}\right)U\right]=\\
G\left(p,x\right)V\left(x+\mathbf{i}\hbar\partial_{p}/2\right)\label{eq:VG-exact-2}
\end{multline}
for the Dyson equation of the second variable $x_{2}$. Above, the
potential $V\left(x\right)\equiv V\left(\boldsymbol{x},t\right)$
could be time-dependent. The substitution $x\pm z/2\rightarrow x\pm\mathbf{i}\hbar\partial_{p}$
is sometimes called a Bopp's rule \cite{Polkovnikov2009}. Note the
difference between the transformation rules \prettyref{eq:VG-exact}
and \prettyref{eq:G-Wigner-UMGU}, when in the later the potential
does not commute with the parallel shift $U$, whereas $V\left(x\right)$
commutes with $U$ in \prettyref{eq:VG-exact}.

Thanks to their generality, expressions \prettyref{eq:D-Wigner-1},
\prettyref{eq:D-Wigner-2} \prettyref{eq:DD_Wigner_transform}, \prettyref{eq:VG-exact}
and \prettyref{eq:VG-exact-2} can be used to transform any Dyson
equation for a two-point Green function $G\left(x_{1},x_{2}\right)$
in the real space into an equation for the associated quasi-classical
Green function $G\left(p,x\right)$ in the phase-space. Nevertheless,
these relations are of purely formal interest, since the expressions
for $\mathfrak{F}_{s}\left(x,\mathbf{i}\hbar\partial_{p}\right)$
sounds hardly to be found in any concrete case. Even for a given gauge-potential
$A_{\mu}$, it is difficult to believe one will be able to find a
compact expression for the associated parallel transport operator
$U\left(x_{1},x_{2}\right)$. Anyways, we do not really need the full
expressions, and we would be already happy to find a systematic expansion
of the relations \prettyref{eq:D-Wigner-1}, \prettyref{eq:DD_Wigner_transform}
and \prettyref{eq:VG-exact}. This would indeed induce a systematic
expansion of the equation for $G\left(p,x\right)$. Such a natural
expansion is provided by the parameter $\hbar$, and it is called
a quasi-classical expansion \cite{Hillery1984}. For instance we have,
expanding \prettyref{eq:VG-exact} at first order
\begin{multline}
V\left(x-\mathbf{i}\hbar\partial_{p}/2\right)G\left(p,x\right)=V\left(x\right)G\left(p,x\right)\\
+\mathbf{i}\dfrac{\hbar}{2}\dfrac{\partial V}{\partial x^{\mu}}\dfrac{\partial G}{\partial p_{\mu}}+\mathcal{O}\left(\left(\dfrac{\hbar}{\tilde{p}\tilde{x}}\right)^{2}\right)\label{eq:VG-expansion}
\end{multline}
and
\begin{multline}
G\left(p,x\right)V\left(x+\mathbf{i}\hbar\partial_{p}/2\right)=G\left(p,x\right)V\left(x\right)\\
-\mathbf{i}\dfrac{\hbar}{2}\dfrac{\partial G}{\partial p_{\mu}}\dfrac{\partial V}{\partial x^{\mu}}+\mathcal{O}\left(\left(\dfrac{\hbar}{\tilde{p}\tilde{x}}\right)^{2}\right)
\end{multline}
 here $\hbar/\tilde{p}\tilde{x}$ represents a dimensionless phase-space
volume, when $\tilde{x}$ and $\tilde{p}$ are characteristic momentum
and position of the system. In other words, the characteristic phase-space
extension of the system $\tilde{p}\tilde{x}$ should be larger than
$\hbar$ for the above expansions to be valid. 

In the following we keep only terms linear in $\hbar$, which has
to be understood as expansion in $\hbar/\tilde{p}\tilde{x}$. Also,
to avoid the calculation of irrelevant terms, we expand the Wigner
transformation of $\hbar D_{\nu}G$ instead of $D_{\nu}G$, since
the covariant derivatives always appear with $\hbar$ factor in the
equation of motion. One has then 
\begin{multline}
\hbar\int dz\left[e^{-\mathbf{i}p\cdot z/\hbar}U\left[D_{\nu}\left(x_{1}\right)G\left(x_{1},x_{2}\right)\right]U\right]=\\
\hbar\left(\dfrac{1}{2}\mathfrak{D}_{\nu}\left(x\right)-\mathbf{i}\dfrac{p_{\nu}}{\hbar}\right)G\left(p,x\right)\\
+\dfrac{\hbar}{2}\left(\dfrac{3}{4}F_{\mu\nu}\left(x\right)\partial_{p}^{\mu}G+\dfrac{1}{4}\partial_{p}^{\mu}G\left(p,x\right)F_{\mu\nu}\right)\label{eq:DG-expand}
\end{multline}
(we drop the variables for $U\left(x,x_{1}\right)$ and $U\left(x_{2},x\right)$
in the following, since there is no more possible confusion) and 
\begin{multline}
\hbar\int dz\left[e^{-\mathbf{i}p\cdot z/\hbar}U\left[G\left(x_{1},x_{2}\right)D_{\nu}^{\dagger}\left(x_{2}\right)\right]U\right]\\
=\hbar\left(\dfrac{1}{2}\mathfrak{D}_{\nu}\left(x\right)+\mathbf{i}\dfrac{p_{\nu}}{\hbar}\right)G\left(p,x\right)\\
+\dfrac{\hbar}{2}\left(\dfrac{1}{4}F_{\mu\nu}\left(x\right)\partial_{p}^{\mu}G+\dfrac{3}{4}\partial_{p}^{\mu}G\left(p,x\right)F_{\mu\nu}\right)\label{eq:GD-expand}
\end{multline}
for the single covariant derivative. For the second order derivative,
only the four first lines of \prettyref{eq:DD_Wigner_transform} are
kept, since the other ones are of higher order ; even the term $\int ds\mathscr{\left[\mathfrak{F}\right]}\mathfrak{D}G$
are of $\hbar^{2}$ order in the expansion of the Wigner transformation
of $\hbar^{2}D_{\mu}D_{\mu}G$ and its adjoint. We then get 
\begin{multline}
\hbar^{2}\int dz\left[e^{-\mathbf{i}p\cdot z/\hbar}U\left[D_{\nu}D^{\nu}\left(x_{1}\right)G\left(x_{1},x_{2}\right)\right]U\right]\\
=-\mathbf{i}\hbar p_{\nu}\mathfrak{D}^{\nu}\left(x\right)G\left(p,x\right)-p_{\nu}p^{\nu}G\left(p,x\right)\\
-\mathbf{i}\hbar p^{\nu}\left\{ \dfrac{3}{4}F_{\mu\nu}\left(x\right)\partial_{p}^{\mu}G+\dfrac{1}{4}\partial_{p}^{\mu}G\left(p,x\right)F_{\mu\nu}\right\} \label{eq:DDG-expand}
\end{multline}
and 
\begin{multline}
\hbar^{2}\int dz\left[e^{-\mathbf{i}p\cdot z/\hbar}U\left[G\left(x_{1},x_{2}\right)D_{\nu}^{\dagger}D^{\nu\dagger}\left(x_{2}\right)\right]U\right]\\
=+\mathbf{i}\hbar p_{\nu}\mathfrak{D}^{\nu}\left(x\right)G\left(p,x\right)-p_{\nu}p^{\nu}G\left(p,x\right)\\
+\mathbf{i}\hbar p^{\nu}\left\{ \dfrac{1}{4}F_{\mu\nu}\left(x\right)\partial_{p}^{\mu}G+\dfrac{3}{4}\partial_{p}^{\mu}G\left(p,x\right)F_{\mu\nu}\right\} \label{eq:GDD-expand}
\end{multline}
which complete the set of correspondence rules. In principle, they
allow to rewrite any Dyson equation into an equation of motion for
the quasi-classical Green function $G\left(p,x\right)$ truncated
at the quasi-classical order. Nevertheless, one usually sums and subtracts
different Dyson equations to obtain transport-like equations, as it
is done in the main text (see also \cite{Kadanoff1962} and \cite{b.kopnin}
for similar treatment, or the literature cited in \prettyref{sec:Transport-equations}).
Note that the above equations can be adapted to any gauge-field, since
the only assumption we have done in this appendix is the definition
of the covariant derivative $D_{\mu}=\partial_{\mu}+\mathbf{i}A_{\mu}$.
Since we treated the generic non-Abelian case, any gauge-theory can
be transformed using the rules above. This is what we do in the main
text, defining a particle-hole $\otimes$ spin $\otimes$ charge gauge-theory.
In the simpler limit of an Abelian gauge theory, when the $A_{\mu}$
components commute among themselves, $G\left(p,x\right)$ and its
derivatives commute with the gauge-field $F_{\mu\nu}\left(x\right)$,
and the covariant derivative $\mathfrak{D}_{\mu}\left(x\right)G\left(p,x\right)$
is just a usual derivative $\partial_{\mu}G\left(p,x\right)$.

\bibliographystyle{apsrev4-1}
\bibliography{/Users/paradis/Desktop/library}

\end{document}